# Accelerated development of amorphous InZnO thin films as transparent conductive Cu diffusion barriers

Stefanie Frick[1], Reyu Sakakibara[2], Manuel Kober-Czerny[1], Arnold Müller[3], Miloš Baljozović[1,4], Franz-Josef Haug[2], Sebastian Siol[1*]

[1] Empa, Swiss Federal Laboratories for Materials Science and Technology, Dübendorf, Switzerland

[2] Institute of Microengineering (IMT), Photovoltaics and Thin-Film Electronics Laboratory, École Polytechnique Fédérale de Lausanne (EPFL), Neuchâtel, Switzerland

[3] Laboratory of Ion Beam Physics, ETH Zurich, Zürich, Switzerland

[4] Department of Chemistry, Marburg University, Marburg, Germany

** Corresponding author:*

*Sebastian Siol, Sebastian.Siol@empa.ch*

**Abstract**

In light of the increasing supply chain concerns regarding silver for solar cell metallization, the replacement of the silver contacts by copper is desirable. As copper diffuses readily in silicon, deposition of an additional diffusion barrier to protect the respective absorber material stacks is required. We investigate multifunctional layers of transparent conductive oxides (TCOs) from the In-Zn-O system to serve as front electrode and Cu diffusion barrier coating, focusing on purely amorphous layers without grain boundaries to impede copper diffusion. We employ a 2D combinatorial approach to simultaneously screen the Zn/(In+Zn) ratio and the oxygen content in a single materials library deposited by magnetron sputtering without intentional substrate heating. Cu diffusion barrier performance was evaluated by depositing Cu on top of intentionally ultrathin In-Zn-O libraries of 7±1 nm on silicon wafers and annealing them at temperatures of 200-450°C. Both the formation of copper silicides, as well as the silicon photoluminescence signal were monitored. The first was detected only after the crystallization of the In-Zn-O films and required annealing temperatures of 450°C and above. Even for extended dwell times of 20 h at a relevant process temperature of 200°C, we find no evidence of Cu ingress for most of our fabricated In-Zn-O compositions, whereas Si|Cu stacks without In-Zn-O barriers showed a reduction of their photo-luminescence intensity already after less than 1 h. These results suggest thin amorphous In-Zn-O films with an optimal Zn/(In+Zn) ratio of ~0.12 and intermediate oxygen deficiency as effective transparent conductive Cu diffusion barriers for solar cell applications.

**Impact statement**

This work provides insights into the compositional dependencies of the In-Zn-O system on the applicability as multifunctional transparent conductive Cu diffusion barrier by introducing an innovative combinatorial approach to significantly accelerate TCO screening.

# 1. Introduction

Photovoltaics (PV) is one of the key technologies to meet the global energy demands though renewable sources with projections indicating a required increase surpassing a factor of 10 in the next decades [1]. In state-of-the art solar cells, silver currently dominates as metal contact. However, silver has been identified as a critical raw material for the future growth in photovoltaic installations. While already more than 10% of the global supply of silver in 2020 was employed in the metallization of solar cells [2], its annual demand is projected to increase to more than 20% in 2027 [3] and above 50% in 2050 [4]. It is therefore reasonable to expect a strong competition between PV and other crucial applications, including medicine, for the available silver resources. Next to research concerning the optimization of the silver metallization process to reduce the required amount per module, the substitution of silver by copper is a promising approach to mitigate this supply chain risk. The utilization of copper entails a challenge due to its highly diffusive nature [5], which necessitates a diffusion barrier between the copper metal contact and the PV absorber to prevent device degradation e.g., by forming deep acceptor states [6]. In order to reduce the number of processing steps and the associated costs, it would be optimal if the transparent conductive electrode, which is required in many solar cell technologies, could additionally serve as a Cu diffusion barrier. A few reports show selected compositions of transparent conductive oxides (TCO) films used as Cu diffusion barriers, e.g., indium tin oxide (ITO) [7–10], tungsten doped indium oxide (IWO) [11,12], aluminum doped zinc oxide (AZO) [13] and tantalum doped tin oxide (TTO) [14]. However, a systematic study on the dependence of the diffusion barrier performance on the TCO stoichiometry is lacking in the current literature to the best of our knowledge. Amorphous thin films exhibit superior diffusion barrier performances compared to their polycrystalline counterparts due to the absence of grain boundaries, which act as fast diffusion pathways [15,16]. In contrast to covalent materials such as silicon, amorphous oxides with large metal s-orbitals forming the bottom of the conduction band have been shown to exhibit electron mobilities competitive with crystalline TCOs [17,18]. The In-Zn-O (IZO) system has been reported to possess a broad compositional range with a stable amorphous phase [19,20] and additionally exhibits higher crystallization temperatures than amorphous indium tin or gallium oxide [21]. Therefore, the In-Zn-O system has a high potential to be applicable as transparent conductive Cu diffusion barriers.

In this study, we perform a co-optimization of electrical and optical properties along with the Cu diffusion barrier performance of In-Zn-O thin films in the compositional space. While early development of transparent conductive In-Zn-O coatings has been accelerated by combinatorial screening of the Zn/(In+Zn) ratio [19,20], we adapted a recently developed combinatorial approach [22] to enable the screening of cation and oxygen content simultaneously. This approach allowed for the identification of the compositional range with the highest conductivity ($2.5 \cdot 10^3$ S/cm), as well as the oxygen content range necessary for high transmittance on a single materials library. Cu diffusion barrier performance was studied by annealing of thin IZO combinatorial libraries (~7±1 nm) deposited on silicon substrates and capped by Cu pads. The low thickness was intentionally chosen to enable an acceleration of the testing procedure. Monitoring the formation of copper silicides via X-ray diffraction (XRD) required annealing temperatures above the crystallization temperature of the IZO films. More sensitive barrier failure tests on equivalent IZO|Cu stacks on Si(n)|$SiO_x$|poly-Si(n) passivating contact stacks were performed by monitoring the decay of the photoluminescence (PL) signal as a measure for barrier

failure. This method revealed that IZO films with sufficient oxygen content exhibit no indication for barrier failure after 20 h at 200°C, a typical temperature for curing treatments during the subsequent metallization process for Silicon Heterojunction solar cells [23]. Our study demonstrates that IZO thin films are suitable transparent conductive Cu diffusion barriers for solar cell applications over a broad compositional range.

# 2. Materials and methods

## 2.1. Thin film deposition

Combinatorial libraries were deposited in an AJA ATC-1500 sputter tool with a confocal sputter-up geometry. A 2-dimensional combinatorial approach was employed, which involves co-sputtering from four sputter targets simultaneously on a static substrate (i.e., without rotation) at oblique deposition angles as depicted in Fig. 1a. The following targets with 2" diameter were used: In (99.99% purity, HMW Hauner GmbH & Co. KG), Zn (99.995% purity, HMW Hauner GmbH & Co. KG), $In_2O_3$ (99.99% purity, Plasmaterials, bonded) and ZnO (99.99% purity, Plasmaterials, bonded). An open-field magnetic configuration was used for the four unbalanced magnetrons. Metal targets were sputtered with direct current magnetron sputtering (DCMS), while radio-frequency magnetron sputtering (RFMS) was applied to the oxide targets. To obtain the desired compositional spreads, powers of 50 W, 40 W, 10 W and 5 W were set for the $In_2O_3$, ZnO, In and Zn targets respectively. The depositions were performed without intentional heating of the substrates, using a grounded substrate holder and at a pressure of 5.2±0.2 µbar (chamber base pressure before start of deposition $<5\cdot10^{-8}$ mbar). Both gas inlets of Ar (flow of 20 sccm, 99.999% purity) and $O_2$ (flow of 0.25-0.5 sccm, 99.999% purity) were located at the chamber bottom to ensure a homogeneous gas distribution in the deposition chamber. Three different substrate types were used: Corning® EAGLE XG® substrates (50.8 mm x 50.8 mm, 1.1 mm thick) for electrical and optical characterization, as well as commercial silicon wafers (50.8 mm x 50.8 mm, 0.65 mm thick, Micro precision engineering), and passivated silicon wafers for Cu diffusion barrier testing (50.0 mm x 50.0 mm, 0.2 mm thick). The passivated silicon wafer samples consisted of flat crystalline silicon wafers with $SiO_x$|poly-Si(n) passivating contact stacks on both sides. The respective fabrication procedure, as visualized in Fig. S1, is similar to that described in Ref. [24]. Prior to the introduction into the vacuum system, the EXG and the commercial Si substrates were cleaned with acetone and ethanol for each 10 min in an ultrasonic bath. Moreover, an RF sputter cleaning of both substrate types was performed to remove organic adsorbates (3 min at 100V (7-8 W) for EXG substrates) followed by target pre-sputter cleaning in pure Ar (16 min) and target poisoning in the respective deposition atmosphere (4 min). To remove the $SiO_2$ layer of the commercial Si wafer, which itself can act as a diffusion barrier, a more intense RF plasma cleaning of 20 min at 30 W was applied. The deposition time for libraries on EXG substrates was 15 min yielding typical thickness values between 90 and 120 nm, while the deposition time on commercial and passivated Si wafers was decreased to 1 min to reduce the required diffusion length (7±1 nm) and therefore the time required for the barrier testing procedures.

## 2.2. Characterization

The deposited thin film materials libraries were characterized with high-throughput methods on 45 pre-defined sample points.

X-ray fluorescence spectroscopy (XRF) was conducted on a Fischer XRF XDV SDD with 50 kV beam voltage for 3 min per sample to obtain cation composition and film thickness. The thickness determination was calibrated with profilometer measurements (Bruker DektakXT, 2 µm tip radius, 5 mg, 33 µm/s scan speed) on thin film samples of pure $In_2O_3$ and ZnO.

A selected combinatorial library on an EXG substrate was transferred via a mobile UHV transfer cart between the deposition chamber and an X-ray photoelectron spectroscope (XPS PHI Quantera), without breaking the vacuum (highest pressure during transfer was $8 \cdot 10^{-8}$ mbar) to minimize the amount of hydroxide adsorbates which impedes the quantification of the oxygen content in the thin film samples. Survey and core level spectra of In 3d, Zn $2p_{3/2}$ and O 1s as well as the In MNN, Zn LMM and O KLL Auger lines and the valence band region were recorded with monochromatized Al Kα excitation operating at 32 W and 15 kV with a beam diameter of 200 µm in a chamber with pressure below $5 \cdot 10^{-9}$ mbar. Due to the high conductivity of the thin film samples, no charge compensation via neutralizers was applied. An overview of the detailed measurement settings is given in the Supplementary Information (Tab. S1).

To obtain reliable Zn/(In+Zn) and O/(In+Zn+O) ratios as well as H contents, Heavy Ion Time-of-flight elastic recoil detection analysis (HI-ERDA) and Rutherford backscattering spectrometry (RBS) were performed in the Laboratory of Ion Beam Physics at ETH Zurich on selected sample positions [25]. The HI-ERDA measurements employed a 13 MeV $^{127}I$ beam with a total scattering angle of 36° and the RBS measurements a 2 MeV He under a tilt angle of 7°. RBS spectra were simulated with SIMNRA [26] and the HI-ERDA results were analyzed with the Potku software [27].

Sheet resistance measurements were performed on a home-built automated 4-point-probe station equipped with a SP-40045TFQ probe head from Microworld (1 mm probe spacing, 125 µm tip radius, 45 mg pressure) and a Thorlabs LTS300 biaxial stepper motor stage. More details on the setup have been published elsewhere [28]. Sheet resistance values were obtained by a linear fit of the nine current voltage pairs, sweeping from -10 µA to 10 µA with a measurement time of 200 ms per current value.

Hall effect measurements were performed on a Manual Hall Effect System from Microworld (HMS3000-EVM100N2R) for selected material libraries on EXG. A current of 0.5 mA and a magnetic field of 0.580 T were used. Prior to the measurement, 100 nm thick contacts of platinum were deposited onto the IZO libraries, and subsequently, the libraries were cut into 25 pieces with a size of approximately 1 cm x 1 cm.

Transmission and reflection spectra of the materials libraries were recorded in the UV-vis-NIR spectral range (200-1100 nm) in a home-built setup. The setup was equipped with two Ocean Optics HR4Pro XR-ES CCD detectors and an Ocean Optics DH-2000-BAL Deuterium-Halogen light source. Automated mapping at the different sample positions is enabled by the usage of a Thorlabs LTS300 biaxial stepper motor stage, which moves the sample through the < $2 mm^2$ beam path, while the optics remain fixed. More information on the setup has been reported in [29]. An integration time of 10 µs per scan and 18000 scans per sample yielded a total measurement time of 3 min per sample.

Information on the structure of the IZO thin film samples was obtained by X-ray diffraction (XRD) measurements in a Bruker D8 Discover 4-circle goniometer setup equipped with a Cu Kα source and a 2D X-ray detector Eiger (Dectris). The measurements were performed in a 1D mode at 2*Θ* angles between 10° and 90° with an integration time of 0.5 s/step and a step width of 0.05°/step.

## 2.3. Cu diffusion barrier performance

Two different Cu diffusion barrier performance testing methods were investigated, which are illustrated in Fig. S2. In both cases, the samples were annealed in inert atmosphere – to prevent Cu oxidation – at different temperatures and durations. Cu diffusion was monitored in Si|IZO|Cu, Si(n)|SiOx|poly-Si|IZO|Cu, as well as reference Si|Cu and Si(n)|SiOx|poly-Si|Cu stacks using the characterizations described below. Both procedures involved a sequence of annealing steps in inert atmosphere – to prevent Cu oxidation – of the above-described Si|IZO|Cu & Si(n)|$SiO_x$|poly-Si(n)|IZO|Cu libraries, followed by monitoring characterization steps. For both procedures, a reference stack without IZO diffusion barrier (Cu|Si) was tested as well.

The first procedure is based on monitoring copper silicide formation via XRD. The annealing steps were performed in the UHV deposition chamber for 12 h at calibrated temperatures between 300°C and 450°C. As long as no copper silicides were detected in the subsequent XRD measurement, the Si|IZO|Cu stack was annealed at the next higher temperature proceeding in steps of 50°C.

Time-of-flight secondary ion mass spectrometry (ToF-SIMS) was performed using a ToF-SIMS 5 system from IONTOF GmBH. Si|IZO|Cu stacks with three selected IZO compositions after the annealing step at 450°C and one sample before annealing were analyzed. Depth profiling was performed using consecutive cycles of sputtering and analysis overlaid at the same area. A 25 kV and 0.4 pA $Bi^+$ primary ion beam rastered over 200 µm x 200 µm was employed for analysis after each $O_2^+$ (1 kV, 273 nA) sputtering cycle rastered over 500 µm x 500 µm.

The second procedure followed a similar scheme, but monitoring the decay in photoluminescence (PL) signal of the passivated silicon wafer, on which the IZO|Cu library stack had been deposited. The annealing was performed on a hot plate in a glove box (<0.5 ppm $O_2$, <0.9 ppm $H_2O$) at 200°C to prevent crystallization of the IZO barrier and avoid Cu oxidation. A homebuilt PL imaging setup was employed to map the PL signal of the wafer: A defocused 808 nm Ostech LASER excited the area of the library, while PL was detected by a silicon-based camera (PIXIS from Princeton Instruments).

To confirm the Si PL signal being responsible for variations observed during PL imaging, PL spectroscopy was performed on a MonoVista CRS+ Raman setup from Spectroscopy & Imaging GmbH on selected Si|IZO|Cu library positions and the Si|Cu reference. A 514 nm Laser as well as an acquisition time of 6 x 10 s was chosen to account for high carrier lifetimes.

## 3. Results and discussion

### 3.1 Approach for orthogonal cation and oxygen gradients

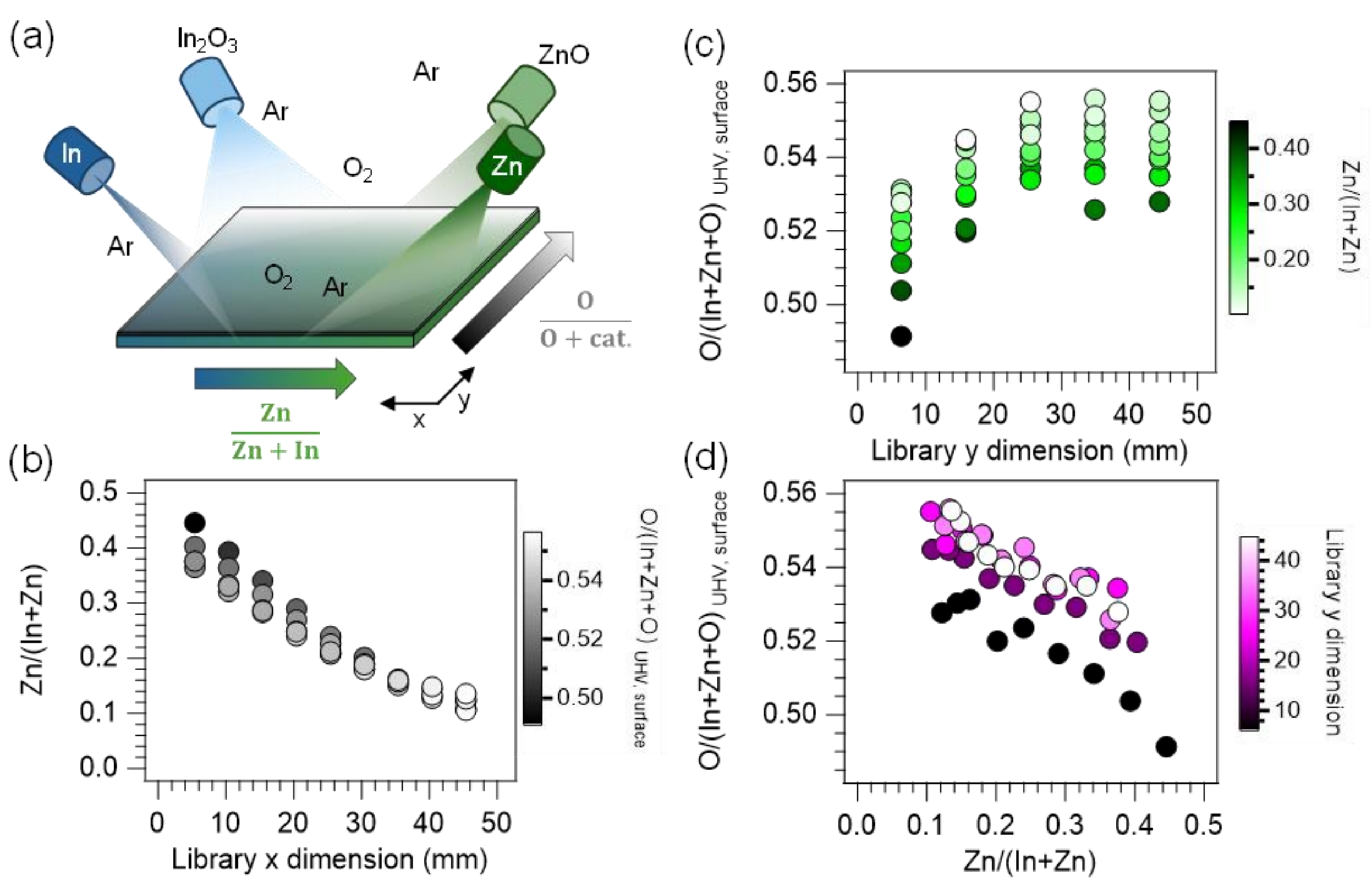

**Fig. 1.** Approach to obtain cation and oxygen content gradients on one combinatorial library: (a) Sketch of the sputter configuration, (b) Zn/(In+Zn) spread calibrated by RBS measurements, (c) surface oxygen composition after a UHV transfer between deposition chamber and XPS as a function of the position on the 50.8 mm x 50.8 mm library and (d) the covered 2D area in composition space.

Typical tests for Cu diffusion barrier performance involve multiple steps of annealing and structural and/or electrical characterization, which is inherently time consuming. Since both cation composition and oxygen stoichiometry critically influence the local and medium-range structure of amorphous TCOs – and consequently their electrical and optical properties – [21,30,31], a large number of samples with various stoichiometries has to be tested. To accelerate this process, combinatorial magnetron sputtering is a valuable tool, since it allows for the preparation of thin films with compositional gradients during a single deposition process. For this work, we adapted a recently developed 2D combinatorial approach for screening simultaneously cation and light anion composition [22], which allowed us to screen a wide range in Zn/(In+Zn) ratio and oxygen content on one library. In **Fig. 1a**, a schematic sketch of the employed magnetron co-sputtering setup is displayed. The symmetric arrangement of the In, Zn, ZnO and $In_2O_3$ targets produce a gradient in Zn/(In+Zn) ratio from left to right, while a gradient of oxygen species arrives at the substrate from the bottom to the top of the substrate in the sketch. In previous work from our group on oxynitrides, we found that the largest spread in anion composition is obtained, if no $O_2$ gas is fed into the deposition chamber [22], i.e. when the deposition is performed in a pure Ar atmosphere. Due to the minute non-stoichiometry in oxygen content required to drastically change the electrical and optical properties, combined with the high deposition rates of the metal targets in metallic mode compared to the oxide targets, a small flow of $O_2$ gas was necessary to obtain material libraries with sufficient transmittance. A series of libraries with different $O_2$ gas flows are presented in Fig. S3, including the effect on visual appearance and electrical properties.

Several characterization methods allow for the quantification of the cation gradient of such obtained libraries. Due to its high speed in acquisition and analysis, XRF was applied to all materials libraries to determine the Zn/(In+Zn) ratio. In order to increase the accuracy of such obtained cation composition, few libraries were investigated with XPS and 15 points on one material library with RBS and ERDA. The corresponding calibration curves can be found in Fig. S4. A typical Zn/(In+Zn) spread achieved on the 2" x 2" libraries is 0.10 – 0.43 (see **Fig. 1b**).

The quantification of the gradient in oxygen content, however, poses a considerable challenge. As visualized in Fig. S5, both RBS and ERDA show the expected reduction of the oxygen content with increasing Zn/(In+Zn) ratio. Moreover, the RBS measurements yield oxygen concentrations close to the stoichiometric composition expected for the respective Zn/(In+Zn) ratios. Contrarily, no significant trend in off-stoichiometry along the y dimension of the library is measurable for neither the current RBS nor ERDA results. Similarly, the quantification of the oxygen gradient of the library using XPS after exposure to air is corroborated by the presence of adsorbates on the surface (see Fig S6 and Fig. S7), which are likely related to hydroxide species (binding energy ~532 eV [32]). Trials of fitting both O 1s components and including only the metal oxide related component led to a correction in the direction of the expected trend. The cleanest gradient in oxygen concentration was, however, obtained by XPS after a UHV transfer between deposition chamber and XPS and is displayed in **Fig. 1c**. While for the lower oxygen contents a steady increase is apparent, it potentially flattens out for higher oxygen contents. The reader's attention may be drawn to the absolute values of the oxygen content which are lower than expected. This has been reported for XPS measurements on UHV transferred Sn doped $In_2O_3$ as well and has been attributed to inaccuracies in the employed sensitivity factors [33]. By combining the obtained Zn/(In+Zn) ratio and oxygen content, **Fig. 1d** visualizes the 2-dimensional composition space covered by the one combinatorial deposition.

While the XPS measurements after air exposure as well as the ERDA and RBS measurements do not enable direct quantification of the gradient in oxygen non-stoichiometry, they may provide indirect indications for the presence of such gradient. As displayed in Fig. S5c, the proportion of the fitted O 1s adsorbate component presents an approximately linear increase in the direction of lower expected oxygen contents. This trend. might relate to a larger number of oxygen deficient surface sites, which can be occupied by hydroxide species. Additionally, the hydrogen content as determined from ERDA increases approximately linearly with decreasing expected oxygen content as well (see Fig. S5d). This gradient might be caused by a less dense structure as reported by Koida *et al.* for preparations with lower oxygen contents [34]. Alternatively, *ab initio* calculations suggest the passivation of undercoordinated In atoms by hydrogen, leading to potentially higher hydrogen contents in case of more oxygen-deficient samples [35,36]. Due to inherently difficult quantification of oxygen off-stoichiometry, the y position on the material libraries is indicated as a proxy for the oxygen content in the following figures.

## 3.2 Optical, electrical and structural characterization

### 3.2.1 Structural screening

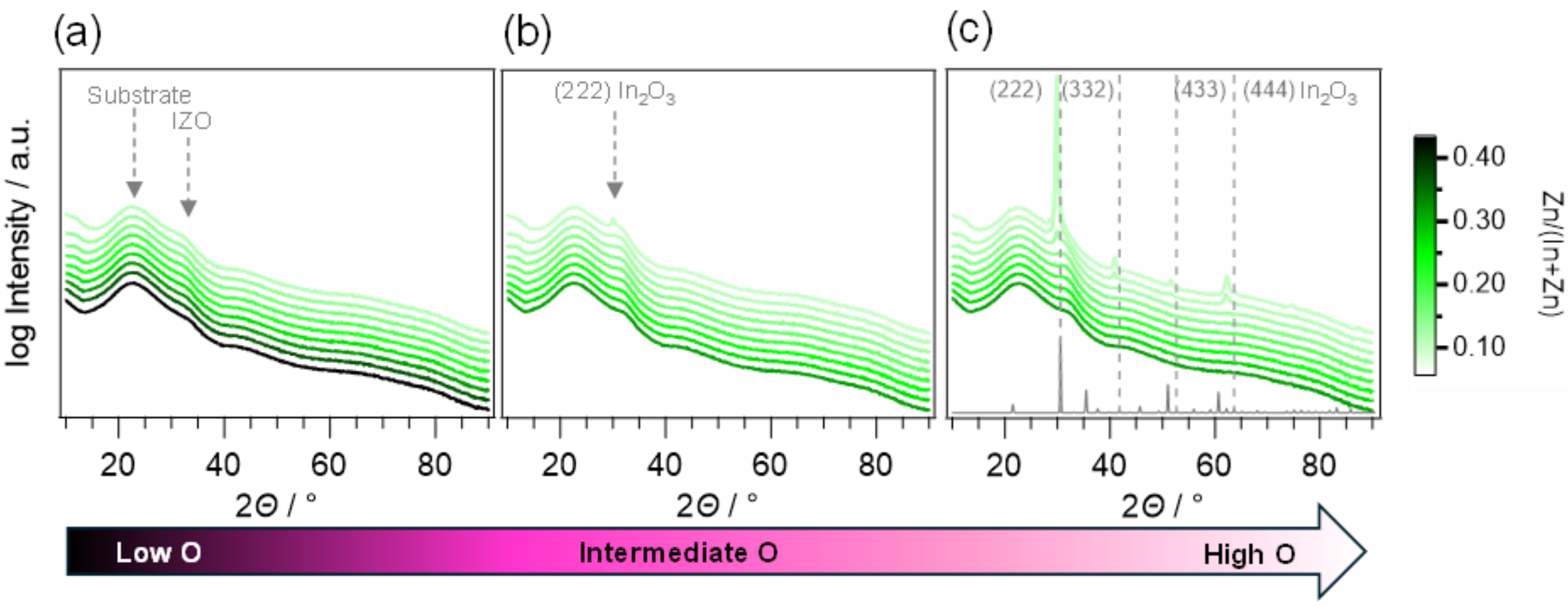

**Fig. 2.** Structural screening of the In-Zn-O-$v_O$ thin film library samples based on X-ray diffraction. The Zn/(In+Zn) ratio is indicated as a color code. The oxygen content of the samples increases from the lowest (a) over intermediate (b) to the highest (c). A reference diffraction pattern from the ICSD database of bixbyite $In_2O_3$ is shown in grey [37].

XRD mapping was performed to investigate the composition dependence of the long-range structure of the IZO libraries. In **Fig. 2**, the diffraction patterns of the samples with different oxygen contents and Zn/(In+Zn) ratios are displayed. Besides the amorphous humps related to the EXG substrate, most of the compositions show a hump at 2$\Theta$ angles around 31-32° as a sole feature, which has been reported in previous studies for X-ray amorphous IZO thin films [19,38]. Nano-crystalline regions with sizes below the detection limit of the XRD mapping may be present as it has been found in X-ray amorphous indium oxide films deposited at ≤ 0°C with PLD [39]. Only the sample positions which combine a low Zn/(In+Zn) ratio as well as a low oxygen deficiency exhibit clear crystalline features attributed to the bixbyite $In_2O_3$ structure. Possessing a preferential (222) orientation, the reflections show a shift towards lower 2$\Theta$ angles compared to pure $In_2O_3$, hence an increased lattice constant. Although this observation appears counterintuitive given the smaller ionic radius of $Zn^{2+}$ compared to $In^{3+}$ [40], it has been reported previously for as deposited IZO thin films [20]. The comparison of the diffractograms in **Fig. 2** indicates that the Zn/(In+Zn) composition, at which the transition between X-ray amorphous and crystalline occurs, depends on the oxygen stoichiometry: For the highest oxygen contents the transition occurs Zn/(In+Zn) ≤ 0.13, for intermediate oxygen contents the onset is shifted below 0.1 and for the lowest oxygen contents crystallization is not observed for any Zn/(In+Zn) ratios covered by the presented library. An additional parameter, which is known to reduce the Zn/(In+Zn) ratio, at which the transition occurs and consequently broaden the amorphous phase space, is low deposition temperature [38].

### 3.2.2 Electrical properties

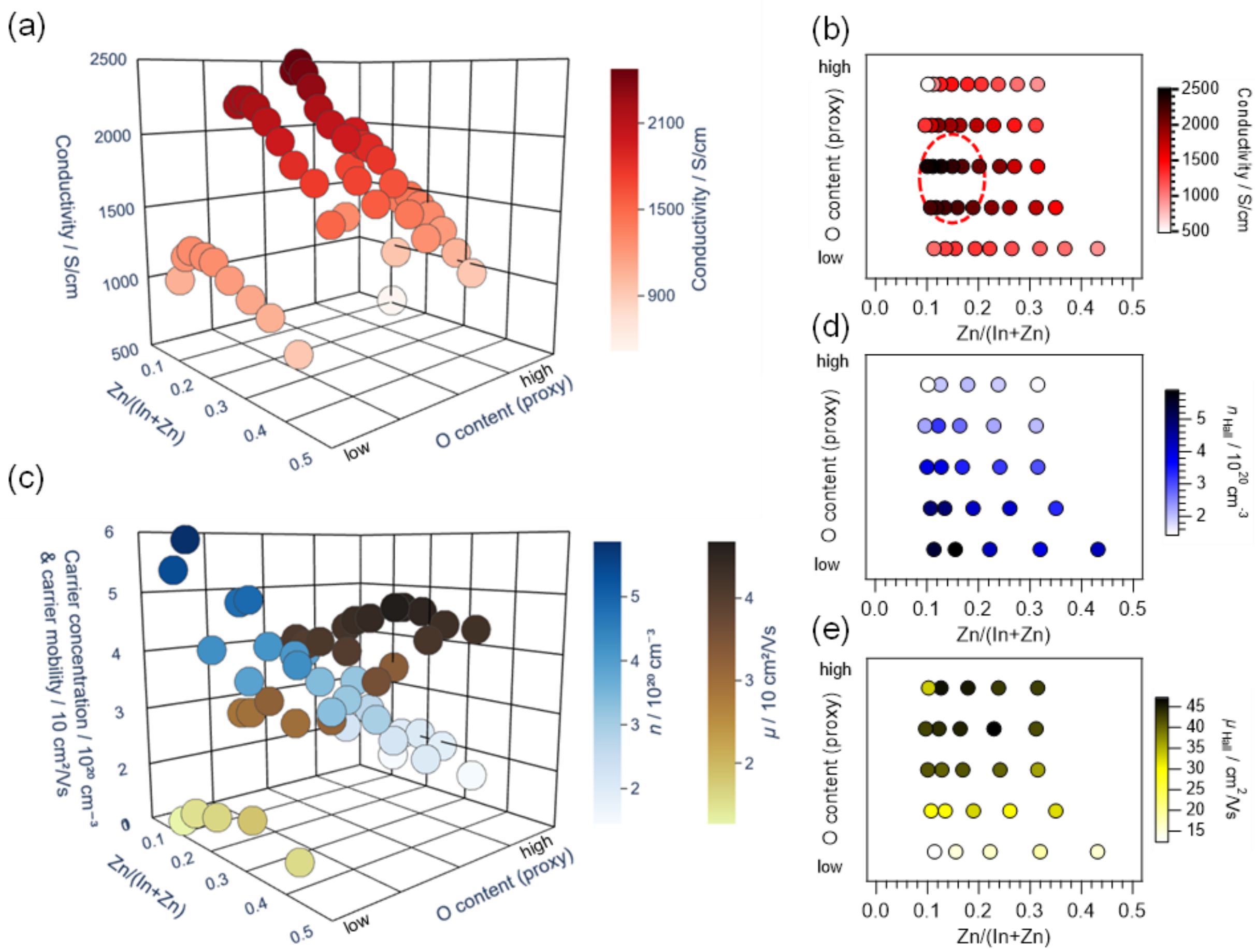

**Fig. 3.** Electrical property screening for the In-Zn-O-$v_O$ thin film library displayed in Fig. 2 based on 4-point-probe measurements for conductivity (a,b) and Hall-effect measurements for charge carrier density *n* and mobility *μ* (c-e). The dashed red region in (a) indicates the compositional regions with conductivities above $2{\cdot}10^3$ S/cm.

The electrical conductivity maps of the IZO library presented in **Fig. 3a,b** showcase the potential of the 2D combinatorial approach for accelerated property optimization. The 2D combinatorial approach allows for the determination of the maximum conductivity achievable for a given set of process parameters in both cation and anion compositional space simultaneously on the same substrate. The conductivity increases steadily with decreasing Zn content reaching a maximum followed by a steep drop for the lowest Zn/(In+Zn) ratio, while the oxygen stoichiometry imposes an additional increase of the conductivity from high to intermediate and a subsequent decrease for low oxygen contents. The superposition of both effects yields a global conductivity maximum of $2.48 \cdot 10^3$ S/cm for a Zn/(In+Zn) ratio of 0.11, while the optimal Zn/(In+Zn) ratio seems to slightly increase for non-optimized oxygen contents. To dive deeper into its origin, Hall effect measurements were performed on the library and presented in **Fig. 3c-e**. Since the Hall effect samples are of the size of ~1x1 $cm^2$, the absolute values are to be understood as averages over a certain spread in cation content and oxygen stoichiometry. The strongest effects on the Hall effect results are observed by the variation of the oxygen content: The concentration of charge carriers shows an approximately linear increase between 1.5 and $5.9 \cdot 10^{20}$ $cm^{-3}$ with increasing oxygen deficiency. Concurrently, the mobility drops steeply for the highest oxygen deficiency (down to 13 $cm^2$/Vs), while exhibiting a plateau for higher oxygen contents of up to 48 $cm^2$/Vs. An analogous trend as well as comparable absolute values have been reported for PLD grown amorphous indium oxide as a function of the oxygen partial pressure and have been compared to *ab initio* molecular dynamics simulations [21]. The results of the latter suggest that oxygen deficiency leads to clusters of undercoordinated indium atoms which resemble oxygen vacancies in crystalline solids and produce shallow donor states increasing the electron density in the conduction band. With increasing oxygen deficiency, highly undercoordinated pairs of indium atoms are present which resemble metallic In-In bonds, and which represent deep defect states leading to increased scattering and therefore reduced carrier mobility. Regarding the influence of the Zn/(In+Zn) ratio, zinc substitution has been shown computationally not to be donors in indium based amorphous oxides, but instead a small decay in mobility has been suggested [30]. However, the reduction of the carrier concentration with increasing Zn/(In+Zn) ratio in this work is more pronounced than the respective change in mobility, similar to Ref. [19]. An additional noteworthy result of the Hall effect measurements is the strong drop in mobility for the sample with the highest oxygen and lowest zinc content, which is likely related to multi-phase scattering in this partially crystalline thin film [21], potentially due to a band offset at the interface between amorphous and crystalline phase [41].

As a last remark on the influence of stoichiometry on the electrical conductivity, the potential effect of hydrogen as detected by ERDA measurement should be discussed. Hydrogen has been demonstrated both experimentally and computationally to act as shallow donors in crystalline $In_2O_3$ [42,43]. However, *ab initio* calculations have predicted that hydrogen itself does not act as a donor in amorphous indium based oxides, but indirectly increases the charge carrier density to a minor extent by passivating undercoordinated In atoms in oxygen deficient structures [36]. As discussed above, the hydrogen content in the samples follows the gradient in oxygen deficiency over the library. Therefore, it might be tempting to relate the linear increase of the charge carrier density to the increase in hydrogen content. Indeed, assuming a density of crystalline $In_2O_3$ (7.12 g/$cm^3$ [37]), a hydrogen concentration between approximately 1.5 and $3.9 \cdot 10^{20}$ atoms/$cm^3$ would be present, which corresponds to a similar order of magnitude as the determined carrier concentrations. Due to the overlaying gradients in oxygen sub-stoichiometry and hydrogen content in the libraries in this work, a disentanglement of both effects on the charge carrier density is consequently not feasible.

### 3.2.3 Optical properties

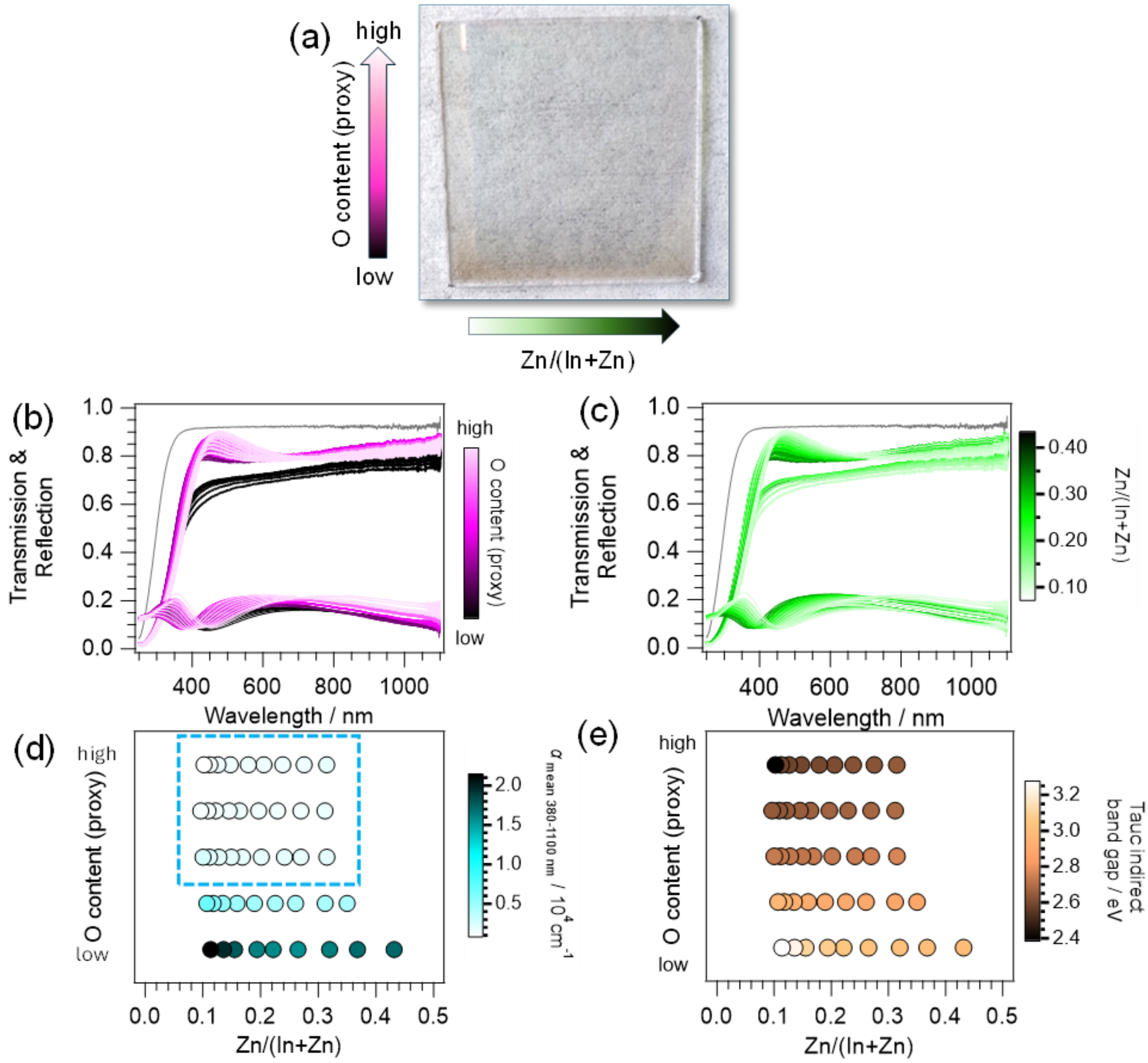

**Fig. 4.** Optical property screening for In-Zn-O-$v_O$ thin film libraries: (a) Visual appearance of the library presented in Fig. 2. UV-vis transmission and reflection spectra of the 45 library samples with color codes based on (b) the oxygen content and (c) the Zn/(In+Zn) ratio. The transmission spectrum of the glass substrate is shown in grey. (d) Absorption coefficient averaged between 380 nm and 1100 nm and (e) optical band gap determined by Tauc analysis with an exponent of $x$ = 2 as a function of Zn/(In+Zn) ratio and oxygen content. The blue dashed region in (d) indicates compositions with averaged absorption coefficients below $5 \cdot 10^3$ $cm^{-1}$.

**Fig. 4a** shows the photograph of a representative material library. On a single substrate, the limit in oxygen deficiency being applicable as TCO can be determined indicated by an increase in absorptance observable as a brown hue of the IZO film. Respective UV-vis transmission and reflection spectra are present in **Fig. 4b,c** as a function of the oxygen content and the Zn/(In+Zn) ratio. The thereof calculated absorption coefficient spectra are displayed in Fig. S8. The spectrally averaged absorption coefficient quantifies the visual impression in a scalar value instrumental for the following co-optimization. While the absorption coefficient is mainly governed by oxygen content, an increase of the Zn/(In+Zn) ratio is accompanied by a slight increase in case of lower oxygen contents and a decrease for higher oxygen contents which exhibit crystalline reflections in XRD. The calculated weighed absorptance as a more device-relevant parameter shows the same trends (for more details see Fig. S9). The parasitic absorption could be related to different phenomena such as absorption or scattering of metallic inclusions [44]. However, neither Zn or In related reflections in XRD were detected nor the development of a component at lower binding energies in the Zn 2p or In 3d spectra were observed in XPS (even for the library transferred in UHV), which would provide proofs for the presence of such metallic inclusions. *Ab initio* molecular dynamics simulations by Medvedeva *et al.* reveal that under-coordinated In atoms in highly oxygen-deficient amorphous indium oxide create deep trap states, which leads to absorption in the visible range [21]. Fig. S10a visualizes the correlation between increased sub-bandgap absorption and decrease in charge carrier mobility, suggesting a common origin of both phenomena.

Bandgap determination in amorphous TCOs has been previously performed by applying the Tauc analysis [45] for direct band gap semiconductors [46,47]. However, the choice of a linear slope region in the plots of $(\alpha h\nu)^{\frac{1}{x}}$ vs. $h\nu$ with $x = \frac{1}{2}$ (see Fig. S11) is not unambiguous. Moreover, the significant sub-bandgap absorption of the most oxygen deficient compositions poses a further challenge to the reliable application of the Tauc analysis. Therefore, four different methods were applied to determine the optical gap from the absorption spectra: Automated Tauc analysis for a (i) direct ($x = \frac{1}{2}$) and (ii) indirect transition ($x = 2$) using the published code from Suram, Newhouse and Gregoire [48], (iii) the maximum in the second derivative of the absorption coefficient indicating the onset of its steep rise, as well as (iv) a set threshold value on the slope of the absorption coefficient. The resulting optical bandgaps for the different methods as a function of the film compositions are compared in Fig. S12. All four methods exhibit an approximately linear decrease in the optical bandgap with increasing oxygen content. However, the subtle influence of the Zn/(In+Zn) varies between the different measures for the optical gap. Therefore, the correlations between the different proxies for the optical gap and the charge carrier density as obtained by Hall effect measurements reproduce a Burnstein-Moss shift [49,50] of the optical gap upon increase of charge carrier density with different quality (see Fig. S10b-e). Moreover, the absolute values exhibit a significant spread from 2.6 eV for the indirect Tauc gap to 3.7 eV for the direct Tauc gap and intermediate energies for the measures of the absorption onset. A comparison with the XPS binding energy difference between valence band and Fermi level revealed the strongest agreement with the indirect Tauc gaps. Therefore, the dependence of the indirect Tauc band gap on the cation and oxygen composition is selected to be displayed in **Fig. 4e**.

## 3.3 Cu diffusion barrier performance

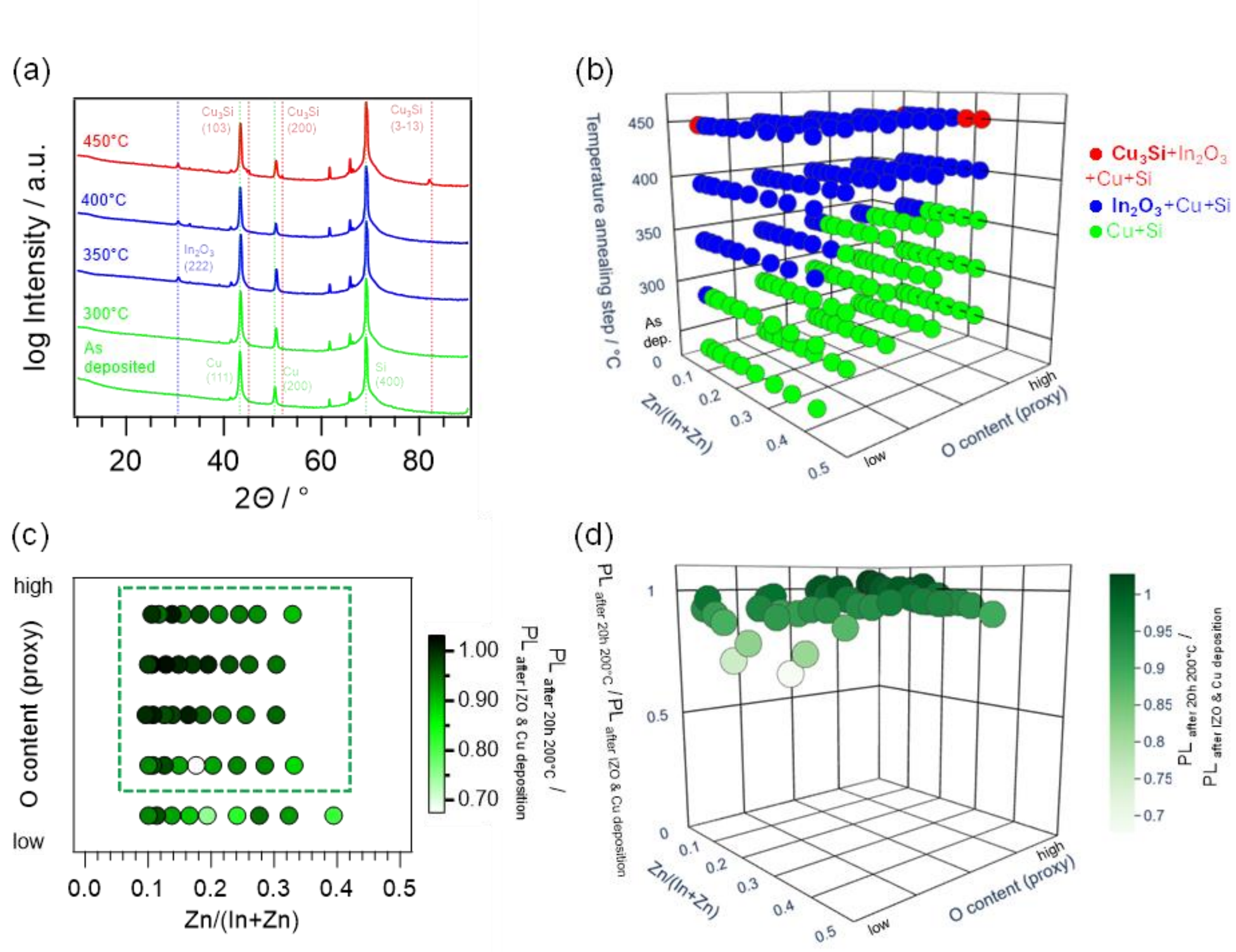

**Fig. 5.** Cu diffusion barrier performance based on silicide formation via XRD (a,b) and ratio of the photoluminescence response of a passivated Si wafer before and after annealing at 200°C for 20 h as a function of Zn/(In+Zn) ratio and oxygen content (c,d). Dashed lines in (a) represent reflections extracted from literature references of fcc-Cu [51], Si [52] (both green), $In_2O_3$ [37] (blue) and $Cu_3Si$ [53] (red).

To simulate accelerated Cu diffusion through the different In-Zn-O compositions for silicon solar cell applications, the In-Zn-O library was replicated in a thinned version (7±1 nm) on a commercial silicon wafer subsequently capped with a raster of 5x9 Cu pads (100 nm) at the respective sampling points (see Fig. S3). The reduced thickness was chosen to decrease diffusion length of Cu through the IZO layer $L$ and therefore testing time $t$ according to $L = \sqrt{Dt}$ ($D$ being the diffusion coefficient). The experimental procedure consisted of annealing steps in inert atmosphere each followed by XRD measurements to monitor copper silicide formation representing the criterion for barrier failure, in analogy to earlier reported Cu diffusion barrier test for different TCO materials [7–14]. In order to potentially reduce the annealing temperatures required to observe barrier failure and therefore improve the comparability to diffusion processes at operation conditions (max. 80°C), longer annealing times (12 h) were chosen compared to most of the tests reported in literature (5-10 min) [7–14]. However, the explored annealing temperatures are still above typical processing and operation conditions to detect effects in the XRD patterns. In **Fig. 5a**, the diffractograms of the subsequent annealing steps of the sample with the highest oxygen content and lowest Zn/(In+Zn) ratio are displayed. Even though this composition showed $In_2O_3$ related reflections for the 15 x thicker sample on glass, only Cu and Si related peaks are observable for the as-deposited 7 nm thin IZO film indicating an amorphous structure. The amorphous state retained for all IZO compositions after 12 h annealing at 300°C, except for the most oxygen deficient and Zn-poor sample. After the following annealing step (additional 12 h at 350°C), approximately half of the IZO thin films crystallize in a bixbyite-related structure. While the preferential crystallization of the Zn-poor samples is expected as Zn is reported to increase the crystallization temperature [21], the crystallization of samples with lower oxygen content at lower temperatures is surprising. After the subsequent annealing step at 400°C, all covered IZO compositions show crystallization. The first $Cu_3Si$ reflections are detected starting from the following annealing step at 450°C for the samples with the highest In- and/or highest oxygen contents. A reference sample of Cu pads directly deposited on an equivalent Si wafer exhibits $Cu_3Si$ reflections already after 2 h annealing at 300 °C as well as the conversion of all Cu as indicated by the vanishing Cu reflections in the respective diffractogram (see Fig. S13) as well as the disappearing cuprous color of the Cu pads. This highlights the effectiveness of the barrier functionality of the IZO films even at thickness as low as 7±1 nm for the intermediate temperature range.

Although this testing approach is widely reported for the evaluation of Cu diffusion barrier performance, two reasons require the critical review of its informativeness in the specific case of this study. On the one hand, all IZO films are crystallized before copper silicides can be detected, which very likely alters the diffusion mechanism in comparison to the amorphous state in which the IZO films are present during operation and processing. In contrast to the amorphous films, the (partially) poly-crystalline films exhibit grain boundaries, which introduce fast diffusion pathways for Cu species. On the other hand, the detection of copper silicides via XRD requires crystals with a size above the detection limit. However, solar cell absorbers and other semiconductors are highly sensitive to even small ingress of aliovalent ions which corrupt their carefully tuned transport properties. More specifically, it has been shown that Cu contamination as low as $6\cdot10^{14}cm^{-3}$ reduces the minority carrier lifetime in silicon by one order of magnitude [54]. To investigate potential Cu ingress into silicon below the detection limit of XRD, ToF-SIMS depth profiles were obtained on selected samples by $O_2^+$ sputtering. Indeed, Fig. S14 and Fig. S15 reveal that the diffusion profiles broaden significantly even before copper silicides are detectable by XRD, indicating potentially earlier barrier failure. Therefore, a complementary Cu diffusion barrier performance test is required, which is, on the one hand, more sensitive with respect to Cu concentration and, on the other hand, remains in a temperature range below the onset of crystallization of the IZO films. Photoluminescence (PL) imaging is a common technique to map the local effective carrier lifetime of silicon wafers [55] and is consequently a promising method to detect Cu ingress [24]. Therefore, a 7±1 nm thin In-Zn-O library was replicated on a passivated silicon wafer, capped with Cu pads and subjected to annealing at 200°C in Ar, a typical curing temperature for metallization of silicon hetero-junction solar cells [23]. PL imaging was performed on the passivated Si wafer, after IZO and Cu deposition as well as after 20 h of annealing (see Fig. S16). In **Fig. 5c,d**, the ratio between the PL signal before and after annealing is displayed as a function of Zn/(In+Zn) ratio and oxygen content. While the oxygen-richer and Zn-poor compositions show no reduction in PL signal, a small decay is observed for samples with the lowest oxygen contents as well as higher Zn contents. Contrarily, the respective reference of Cu pads directly deposited on a passivated Si wafer without IZO diffusion barrier exhibits a reduction down to 0.14±0.02 times the original PL intensity after annealing as short as 1 h at 200°C. This reduction of the Si PL signal after annealing of the reference Si|Cu sample compared to the Si|IZO|Cu library was confirmed by PL spectroscopy on selected library positions (see Fig. S17). Therefore, the results indicate the applicability of nearly all screened compositions excluding the lowest oxygen contents as effective Cu diffusion barriers with thicknesses as thin as 7±1 nm.

## 3.4 Co-optimization

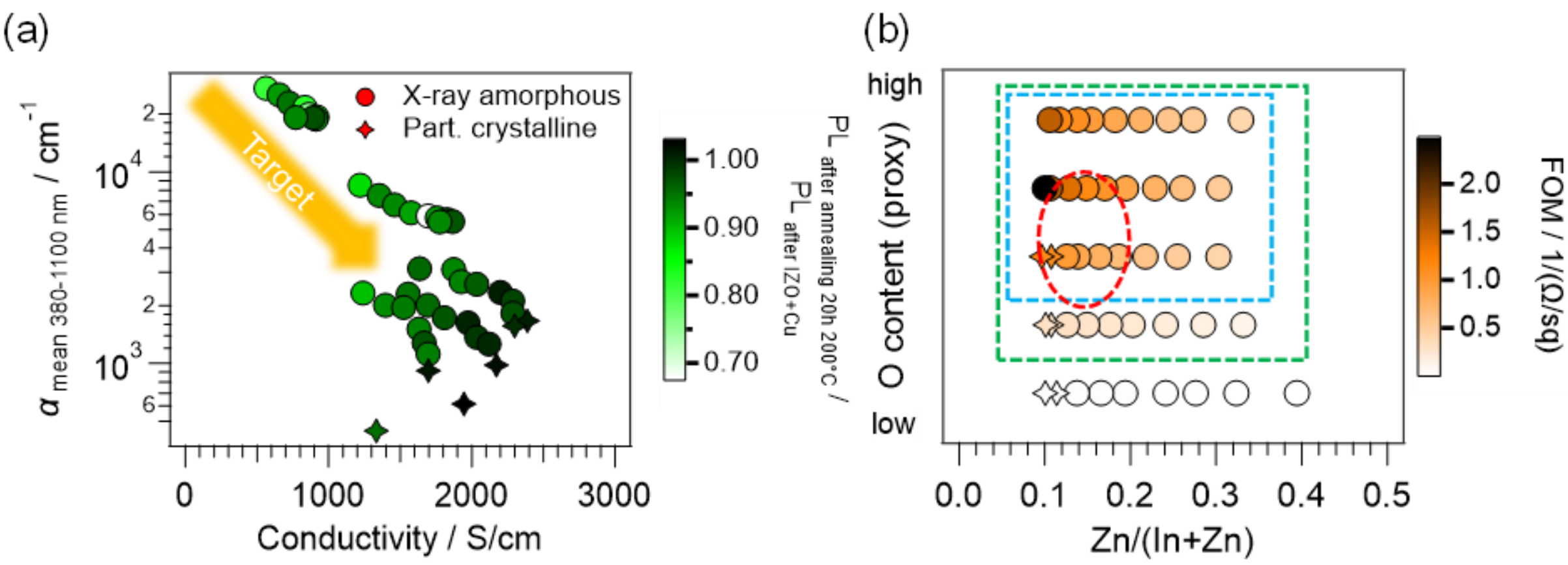

**Fig. 6.** (a) Visualization of the co-optimization of electrical, optical and structural properties combined with the diffusion barrier performance of an In-Zn-O-$v_O$ thin film library. The arrow indicates the targeted optimization direction for TCO materials towards high conductivity and low absorption. Stars indicate the presence of $In_2O_3$ reflections in the XRD pattern of the sample, i.e., partly crystalline samples, while circles represent their absence, i.e., X-ray amorphous samples. (b) Figure of merit as calculated from weighted absorptance and sheet resistance $1/(A_{\mathrm{weighted}}R_s)$ [56] choosing a typical thickness of 100 nm as a function of Zn/(In+Zn) ratio and oxygen content. The dashed lines enclose the compositional regions of interest as determined for the different properties: conductivity (red), absorption coefficient (blue) and Cu diffusion barrier performance (green).

**Fig. 6a** combines the functional properties obtained in the previous sections to enable a visual representation for co-optimization. For a TCO, a trade-off between high electrical conductivity and low absorption coefficient must be established. For the compositional regions with mean absorption coefficients >$3\cdot10^3$ cm$^{-1}$ (i.e., high oxygen deficiency), a lower mean absorption coefficient is accompanied by improved electrical conductivity. Following the discussion above, this can be probably attributed to the presence of deep trap states of under-coordinated metal species, reducing mobility as well as increasing absorption. For samples with intermediate and high oxygen content, the absorption coefficient varies to a smaller extent (note the logarithmic scale in **Fig. 6a**), while the conductivity can be tuned between $1\cdot10^3$ S/cm and $2.5\cdot10^3$ S/cm. The compositions possessing the highest conductivities exhibit mean absorption coefficients around $1\cdot10^3$ S/cm and – notably – are located close to the transition from X-ray amorphous to the onset of X-ray-detectable crystallinity, which confirms literature reports for $In_2O_3$ [39] and In-Zn-O [19]. Moreover, the samples with the best TCO properties – indicated by the color scale in **Fig. 6b** with a figure of merit proposed by Morales-Masis *et al.* [56] – additionally show no indication for diffusion barrier failure. Since UV induced degradation has emerged as a major concern regarding long-term stability of PV-modules [57], we additionally performed accelerated UV-stability testing on an In-Zn-O library (see Supplementary Note 1). We find that the compositional region of interest shows no significant loss in conductivity or reduction in band gap after intense UV exposure equivalent to approximately 6 years of outdoor UV exposure [58] (see Fig. S18). An increase of sub band gap absorption can be mitigated by choosing a Zn/(In+Zn) ratio above the onset of X-ray detectable crystallinity (Zn/(In+Zn) ≥ 0.12). Therefore, the co-optimization suggests the application of the IZO films with Zn/(In+Zn) = 0.12 and intermediate O deficiency to be the optimal choice for the application as multifunctional transparent conductive UV-stable Cu diffusion barrier for temperature sensitive opto-electronic devices.

## 4. Conclusions

In this study, we performed a co-optimization of electrical, optical and barrier properties in the In-Zn-O system for the application as transparent, electrically conductive Cu diffusion barrier. To accelerate this process, we used a 2-dimensional combinatorial approach to simultaneously screen the cation and oxygen content during one TCO deposition process using magnetron sputtering. As verified by XRD, most of the compositions with 0.1 ≤ Zn/(In+Zn) ≤ 0.43 deposited at room temperature exhibit the targeted (X-ray) amorphous structure except for samples with Zn/(In+Zn) ≤ 0.1-0.13 and high oxygen contents showing gradually more intense $In_2O_3$ bixbyite-related reflections. On the same materials library, the optimum in Zn/(In+Zn) and oxygen content maximizing electrical conductivity ($2.5 \cdot 10^3$ S/cm) was determined. Hall measurements explain the observed conductivity trends by revealing the approximately linear increase of charge carrier mobility from 1.5 and $5.9 \cdot 10^{20}$ $cm^{-3}$ along with the gradient in oxygen deficiency and the concurrently steep decay in mobility for the highest oxygen deficiency from 48 to 13 $cm^2$/Vs. Moreover, the limit in oxygen deficiency regarding transmittance indicated by the onset of significant absorption of $\alpha > 1 \cdot 10^4$ $cm^{-1}$ averaged between 380 and 1100 nm was deducible. The comprehensive datasets obtained by the 2D screening approach enable correlations to be drawn among opto-electronic properties, such as absorption in the visible range and charge carrier mobility or optical gap and charge carrier density, as well as potential correlations with the hydrogen content. Diffusion barrier performance tests conducted by extended annealing of Si|IZO|Cu library stacks revealed barrier failures indicated by copper silicide formation after the full crystallization of the IZO library, well above operating and typical post-TCO processing temperatures at which the 7±1 nm thin IZO films remain amorphous. An additional, more sensitive barrier test based on the decay of the photoluminescence signal of a passivated silicon wafer indicates no barrier failure for all compositions with sufficient oxygen content – including those with the optimal TCO properties – for 20 h at 200°C. Accelerated UV-stability testing suggests limiting the Zn/(In+Zn) ratio to ≥ 0.12 to avoid UV induced sub bandgap absorption leading to a maximum conductivity of $2.3 \cdot 10^3$ S/cm accompanied by an averaged absorption coefficient of $2 \cdot 10^3$ $cm^{-1}$ for IZO based Cu diffusion barriers as obtained by deposition without intentional heating or post-annealing. However, the extended range of compositions, which are suitable for the application in the emerging PV concept of Cu metallization, allows for design freedom in fine-tuning the electrical and optical properties for the specific solar cell technology of interest. Additionally, the low process temperatures, at which the amorphous IZO films can be deposited, expand the application field to solar cell technologies with sensitive absorber materials including devices based on amorphous silicon and hybrid halide perovskites.

The study presented additionally showcases the capability of the 2D combinatorial screening approach. Since all sampled compositions are deposited simultaneously, unintentional biases due to variations of chamber and target states in subsequent depositions, e.g., base pressure / contaminations, target poisoning or racetrack, can be avoided and enhance the reliability of drawn conclusions. Moreover, the presented approach is flexible and transferable towards other materials systems. Therefore, this approach might accelerate the development of newly emerging TCO materials and the co-optimization process for various multifunctionalities in the future.

**Acknowledgments**

S. F. acknowledges funding by the Helmut Fischer und Anni Walther Stiftung. Financial support by the ETH ORD program for the development of data management tools is gratefully acknowledged. R.S. and M.K.-C. acknowledges funding from the Swiss National Science Foundation (SNSF) (Project No. 226588). M.B. gratefully acknowledges support by SNSF (Project No. 173720, 212167 and 221265) and the University of Zürich Research Priority Program LightChEC. Kerstin Thorwarth is gratefully acknowledged for technical support. Nathan Rodkey is gratefully acknowledged for fruitful discussions. Romain Carron, Ceren Mitmit and Matteo De Marzi are gratefully acknowledged for providing access to the PLQY and the Hall effect setup of Laboratory for Thin Films and Photovoltaics at Empa.

**Authors' contributions**

**S. F.:** Conceptualization, Investigation, Formal analysis, Visualization, Writing - Original Draft; **R.S.:** Investigation; Writing – Review & Editing; **M.K.-C.:** Investigation; Writing – Review & Editing; **A.M.:** Investigation, Formal analysis; Writing – Review & Editing; **M.B.:** Investigation, Formal Analysis, Writing – Review & Editing; **F.-J.H.:** Investigation; Writing – Review & Editing; **S.S.:** Conceptualization, Supervision, Methodology, Formal analysis, Funding acquisition, Writing – Review & Editing

## 6. Figures

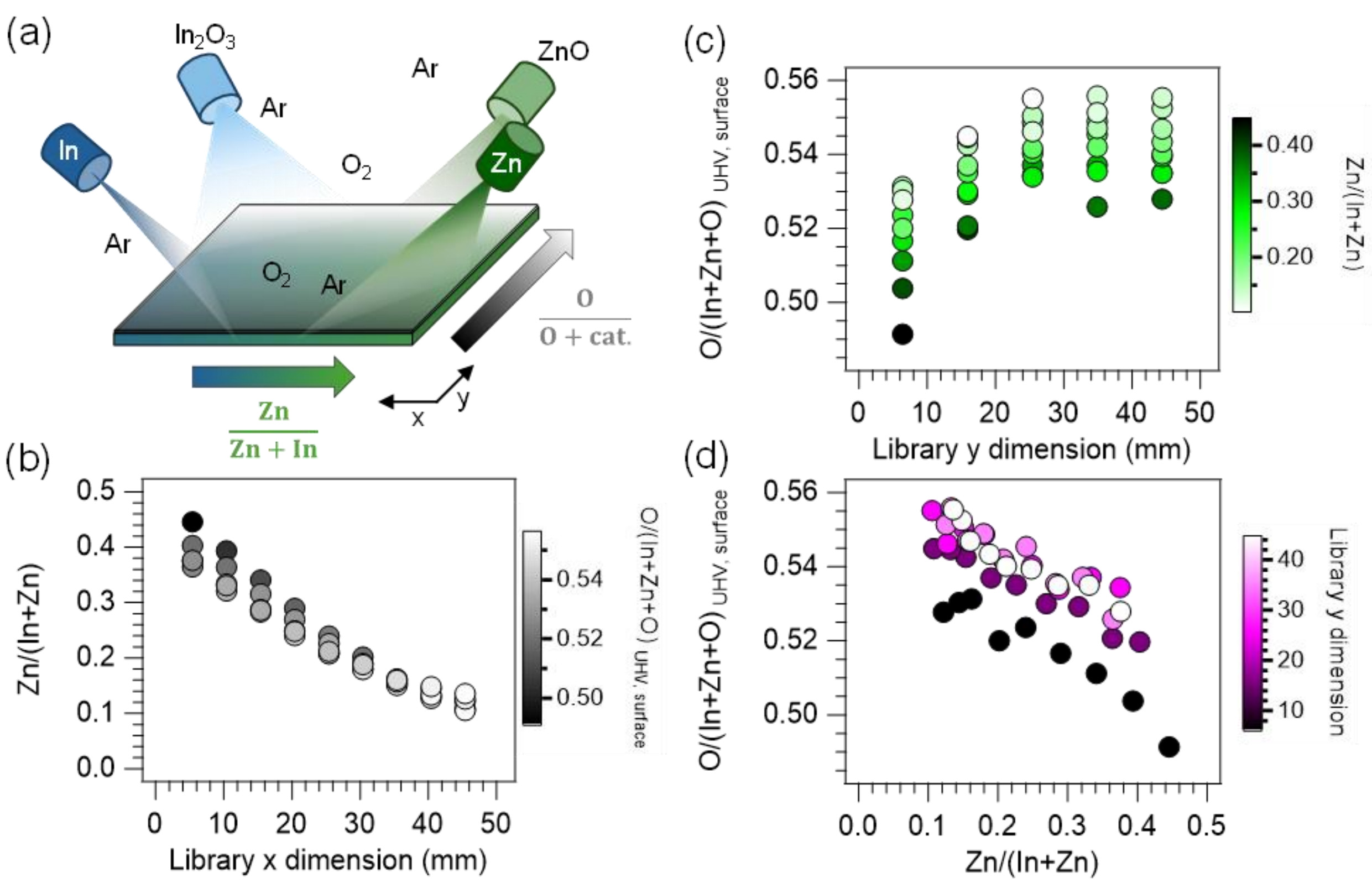

**Fig. 1.** Approach to obtain cation and oxygen content gradients on one combinatorial library: (a) Sketch of the sputter configuration, (b) Zn/(In+Zn) spread calibrated by RBS measurements, (c) surface oxygen composition after a UHV transfer between deposition chamber and XPS as a function of the position on the 50.8 mm x 50.8 mm library and (d) the covered 2D area in composition space.

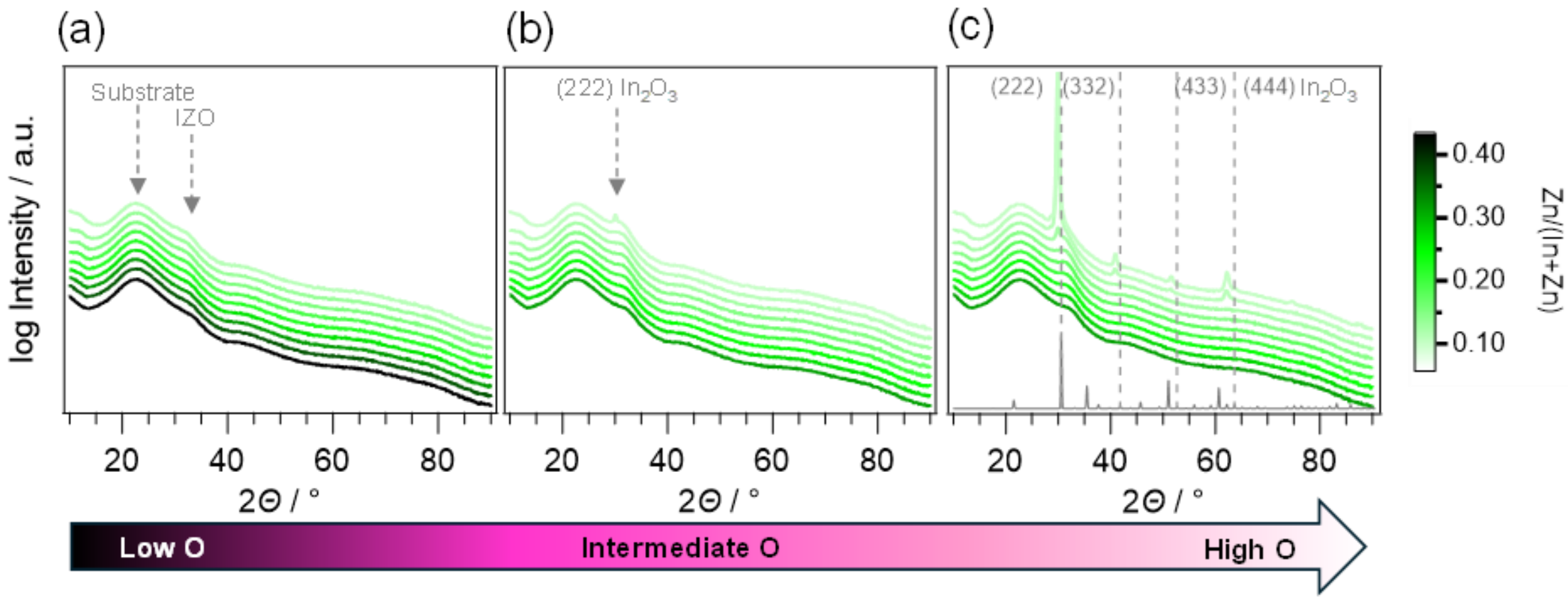

**Fig. 2.** Structural screening of the In-Zn-O-$v_O$ thin film library samples based on X-ray diffraction. The Zn/(In+Zn) ratio is indicated as a color code. The oxygen content of the samples increases from the lowest (a) over intermediate (b) to the highest (c). A reference diffraction pattern from the ICSD database of bixbyite $In_2O_3$ is shown in grey [37].

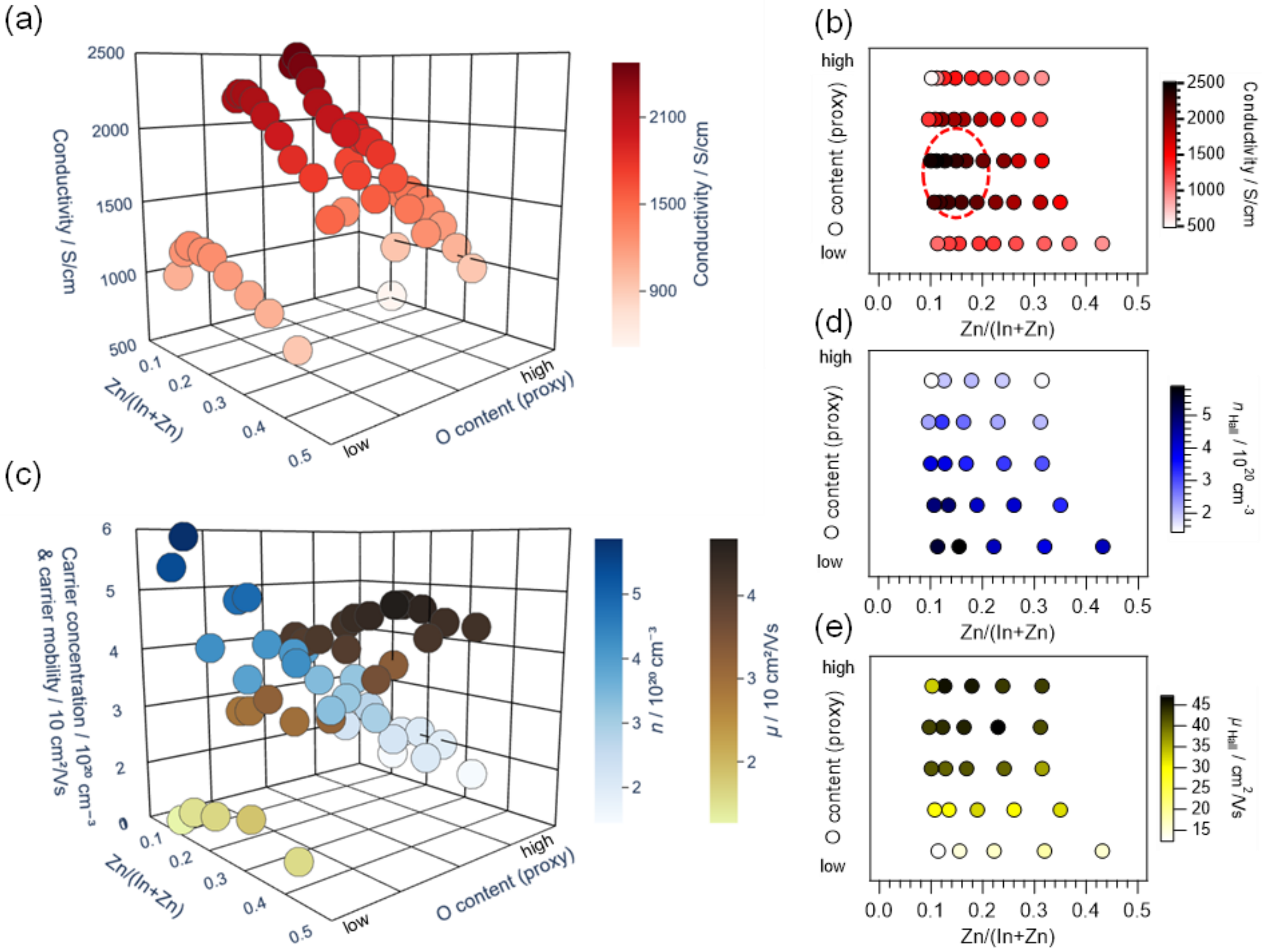

**Fig. 3.** Electrical property screening for the In-Zn-O-$v_O$ thin film library displayed in Fig. 2 based on 4-point-probe measurements for conductivity (a,b) and Hall-effect measurements for charge carrier density $n$ and mobility $\mu$ (c-e). The dashed red region in (a) indicates the compositional regions with conductivities above $2 \cdot 10^3$ S/cm.

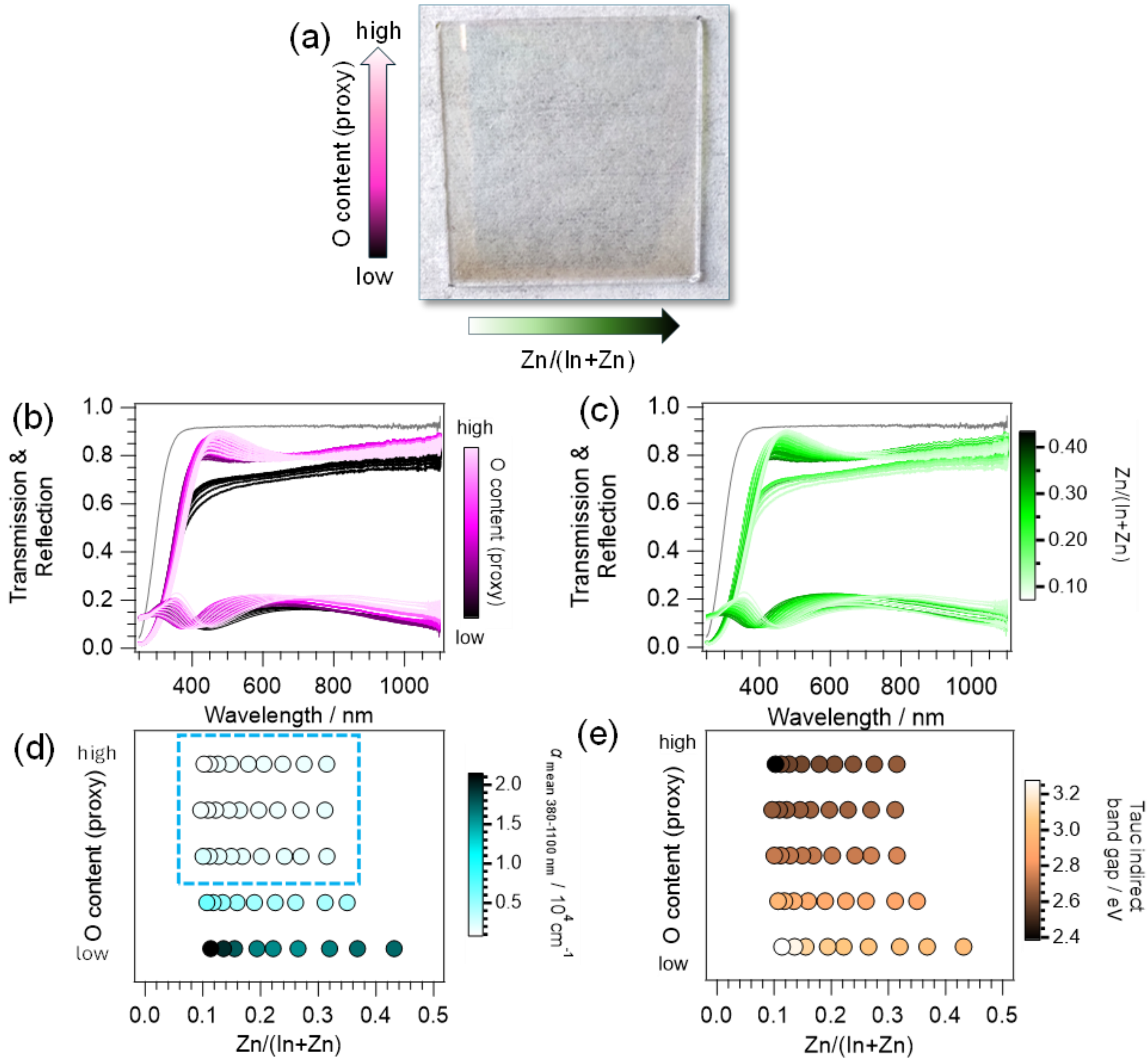

**Fig. 4.** Optical property screening for In-Zn-O-$v_O$ thin film libraries: (a) Visual appearance of the library presented in Fig. 2. UV-vis transmission and reflection spectra of the 45 library samples with color codes based on (b) the oxygen content and (c) the Zn/(In+Zn) ratio. The transmission spectrum of the glass substrate is shown in grey. (d) Absorption coefficient averaged between 380 nm and 1100 nm and (e) optical band gap determined by Tauc analysis with an exponent of $x$ = 2 as a function of Zn/(In+Zn) ratio and oxygen content. The blue dashed region in (d) indicates compositions with averaged absorption coefficients below $5{\cdot}10^3$ $cm^{-1}$.

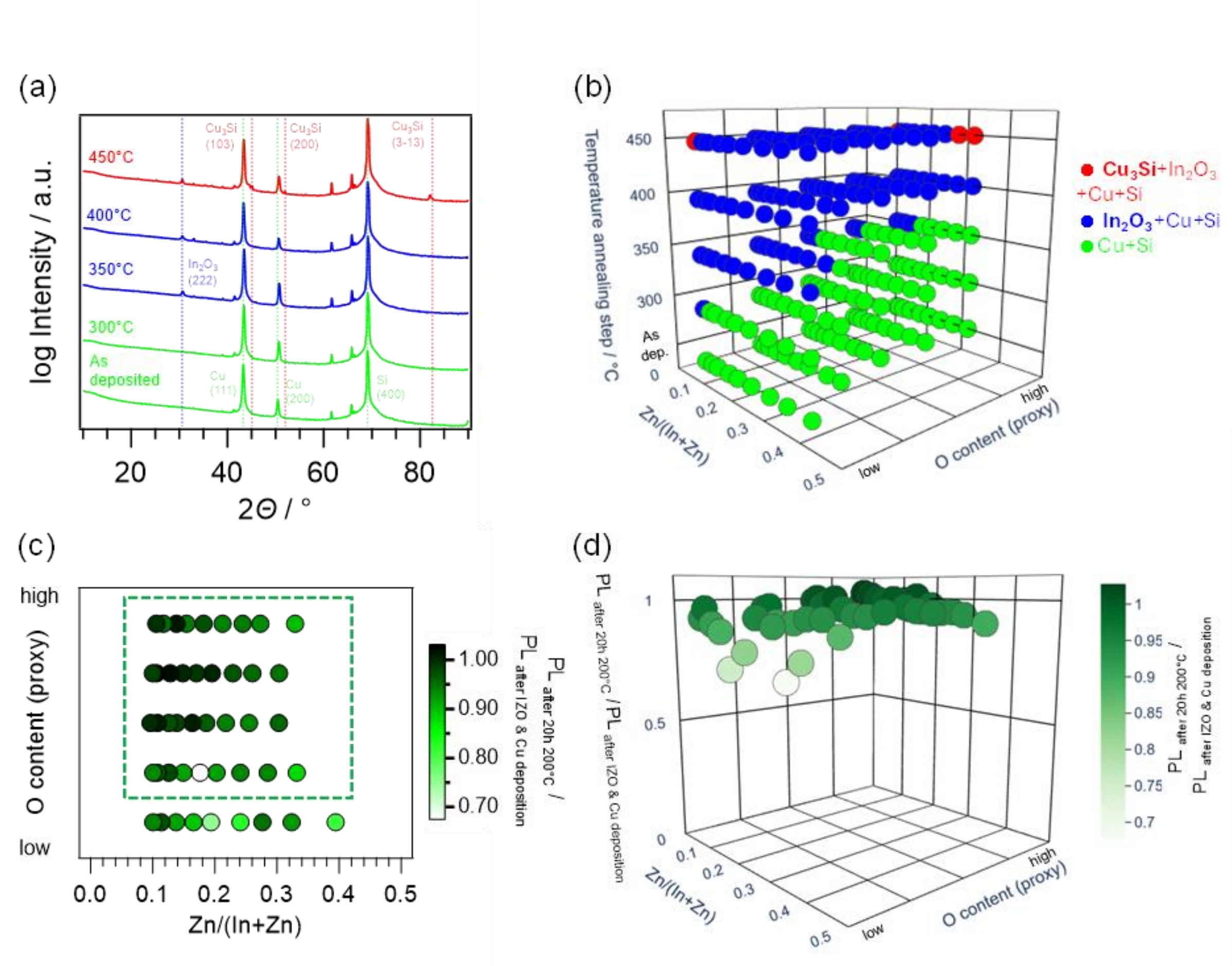

**Fig. 5.** Cu diffusion barrier performance based on silicide formation via XRD (a,b) and ratio of the photoluminescence response of a passivated Si wafer before and after annealing at 200°C for 20 h as a function of Zn/(In+Zn) ratio and oxygen content (c,d). Dashed lines in (a) represent reflections extracted from literature references of fcc-Cu [51], Si [52] (both green), $In_2O_3$ [37] (blue) and $Cu_3Si$ [53] (red).

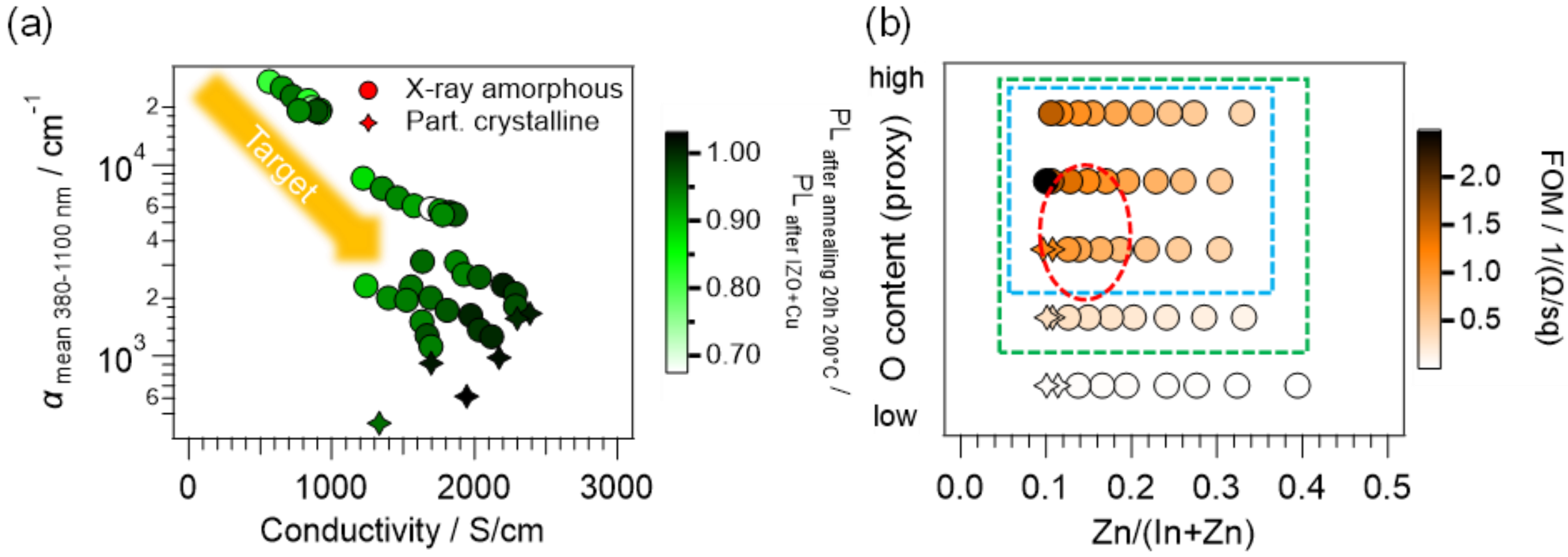

**Fig. 6.** (a) Visualization of the co-optimization of electrical, optical and structural properties combined with the diffusion barrier performance of an In-Zn-O-$v_O$ thin film library. The arrow indicates the targeted optimization direction for TCO materials towards high conductivity and low absorption. Stars indicate the presence of $In_2O_3$ reflections in the XRD pattern of the sample, i.e., partly crystalline samples, while circles represent their absence, i.e., X-ray amorphous samples. (b) Figure of merit as calculated from weighted absorptance and sheet resistance $1/(A_{\text{weighted}}R_s)$ [56] choosing a typical thickness of 100 nm as a function of Zn/(In+Zn) ratio and oxygen content. The dashed lines enclose the compositional regions of interest as determined for the different properties: conductivity (red), absorption coefficient (blue) and Cu diffusion barrier performance (green).

# Supplementary information

# to

# Accelerated development of amorphous InZnO thin films as transparent conductive Cu diffusion barriers

Stefanie Frick[1], Reyu Sakakibara[2], Manuel Kober-Czerny[1], Arnold Müller[3], Miloš Baljozović[1,4], Franz-Josef Haug[2], Sebastian Siol[1*]

[1] Empa, Swiss Federal Laboratories for Materials Science and Technology, Dübendorf, Switzerland

[2] Institute of Microengineering (IMT), Photovoltaics and Thin-Film Electronics Laboratory, École Polytechnique Fédérale de Lausanne (EPFL), Neuchâtel, Switzerland

[3] Laboratory of Ion Beam Physics, ETH Zurich, Zürich, Switzerland

[4] Department of Chemistry, Marburg University, Marburg, Germany

** Corresponding author:*

*Sebastian Siol, Sebastian.Siol@empa.ch*

# 2. Materials and methods

## 2.1 Thin film deposition

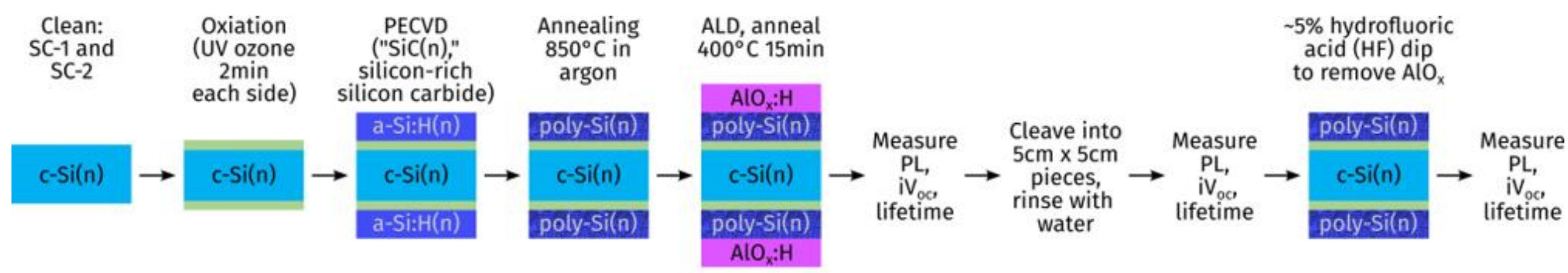

**Figure S1:** Deposition scheme for passivation procedure of a n-type silicon wafer for Cu diffusion barrier testing based on Si photoluminescence.

## 2.2. Characterization

**Table S1:** XPS measurement parameters for an air-exposed library and a UHV-transferred library.

<table>
<tr><th></th><th>Air exposed</th><th>UHV transfer</th></tr>
<tr><td>Beam Power/Voltage</td><td colspan="2">32 W/ 15 kV</td></tr>
<tr><td>Pass energy</td><td colspan="2">55 eV</td></tr>
<tr><td>Step size core level, valence band</td><td>0.1 eV</td><td>0.05 eV</td></tr>
<tr><td>Step size Auger lines</td><td colspan="2">0.01 eV</td></tr>
<tr><td>Pressure</td><td colspan="2">$< 5 \cdot 10^{-9}$ Torr</td></tr>
</table>

## 2.3. Cu diffusion barrier testing

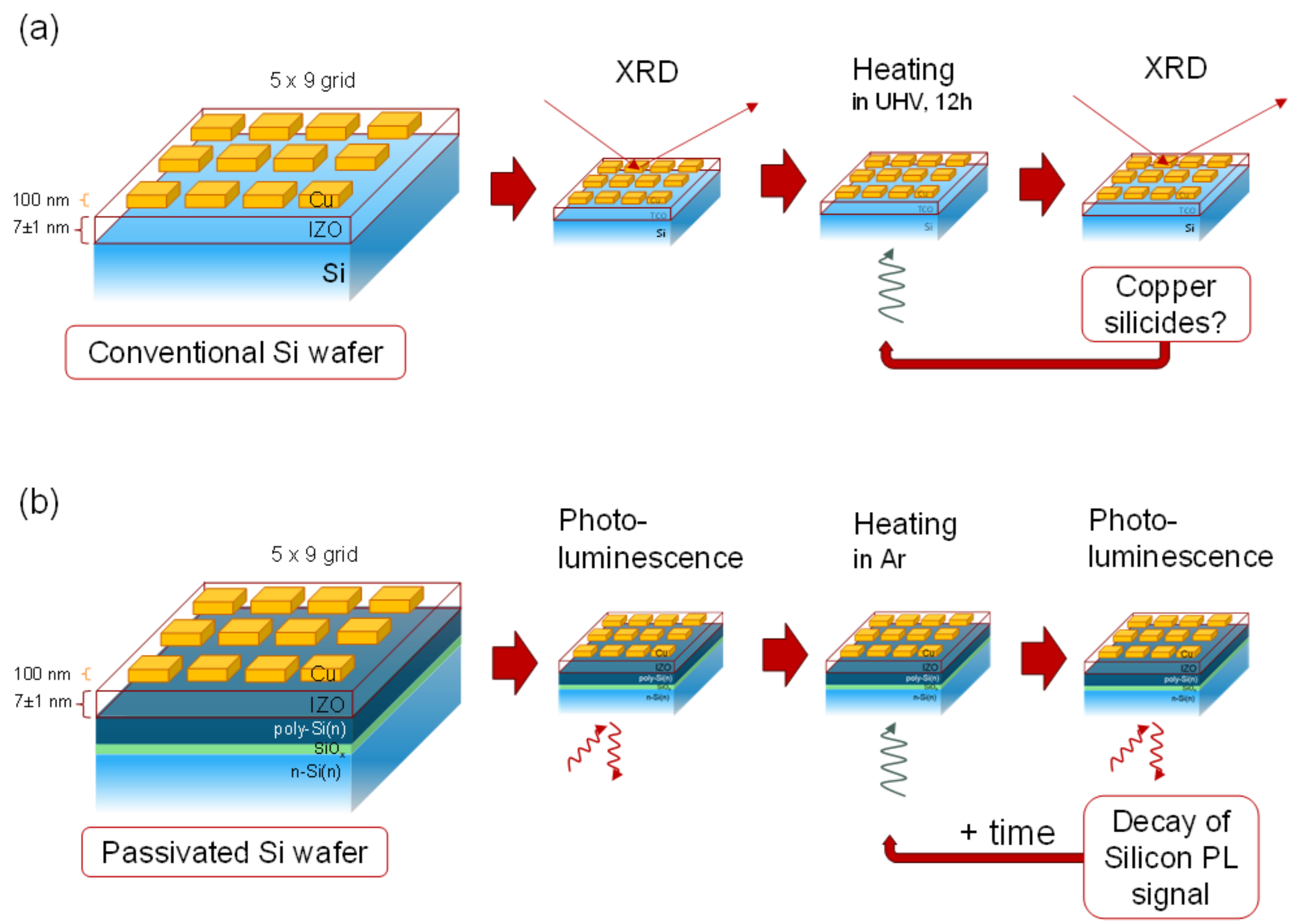

**Figure S2:** Procedures for Cu diffusion barrier testing based on monitoring (a) copper silicide formation via XRD and (b) decay of photoluminescence response of a passivated silicon wafer.

## 3. Results and discussion

### 3.1 Approach for orthogonal cation and oxygen gradients

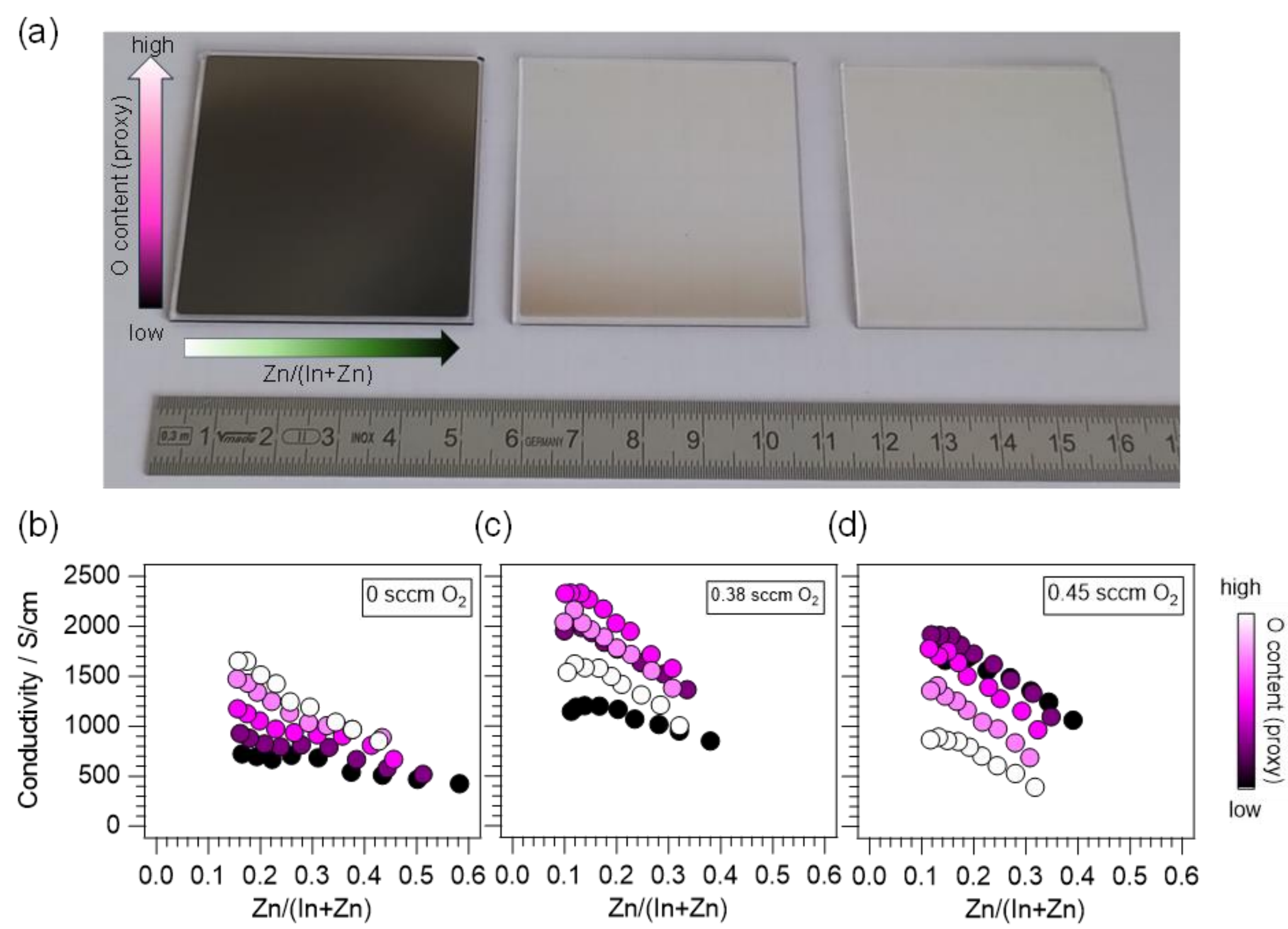

**Figure S3:** Influence of the $O_2$ gas flow on (a) transparency and (b-c) electrical conductivity of In-Zn-O-$v_O$ libraries: (a) Photograph of libraries deposited with 0, 0.38 and 0.45 sccm $O_2$ (from left to right); (b-d) respective achieved electrical conductivity as a function of Zn/(In+Zn) ratio and library y dimension as a proxy for the oxygen content.

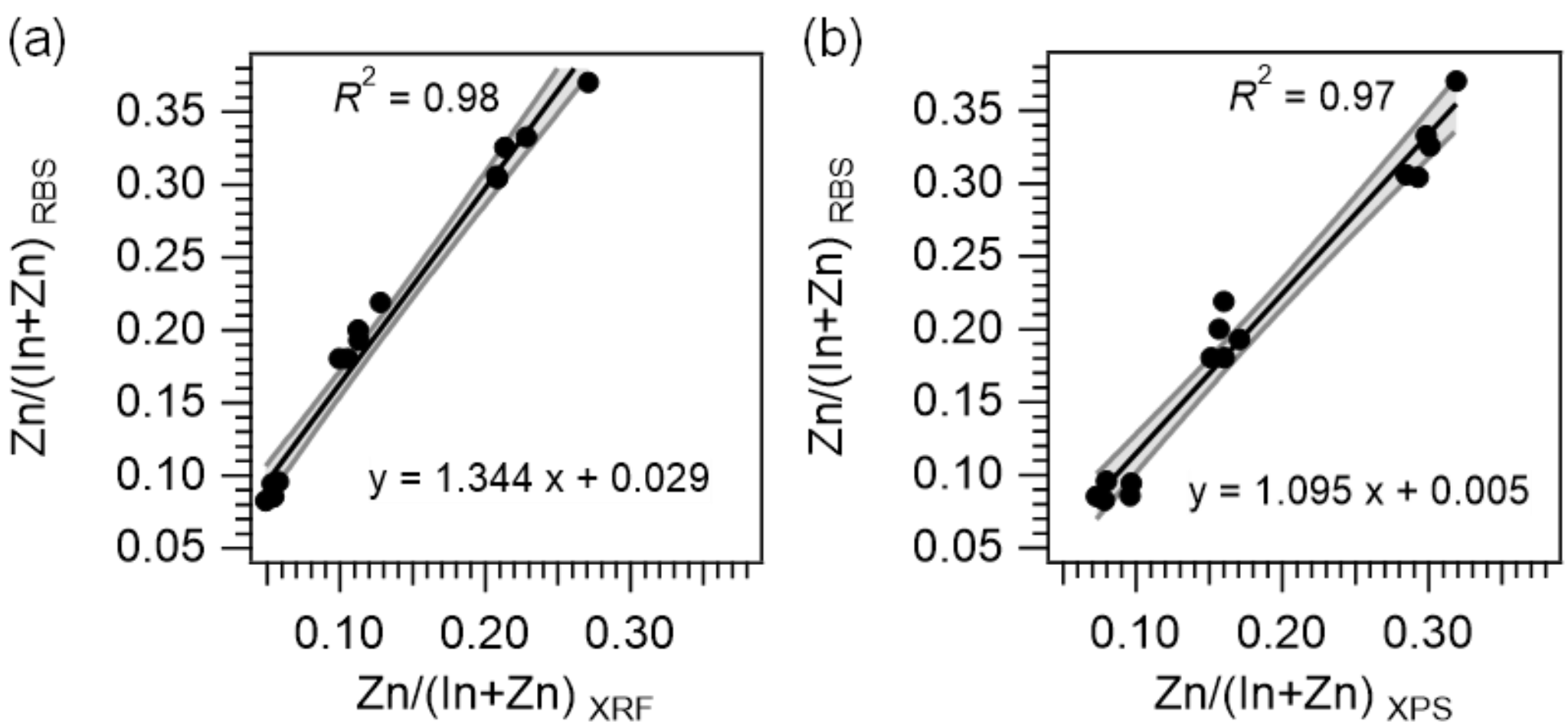

**Figure S4:** Calibration of Zn/(In+Zn) ratio obtained by (a) XRF (b) and XPS (air exposed) with RBS measurements.

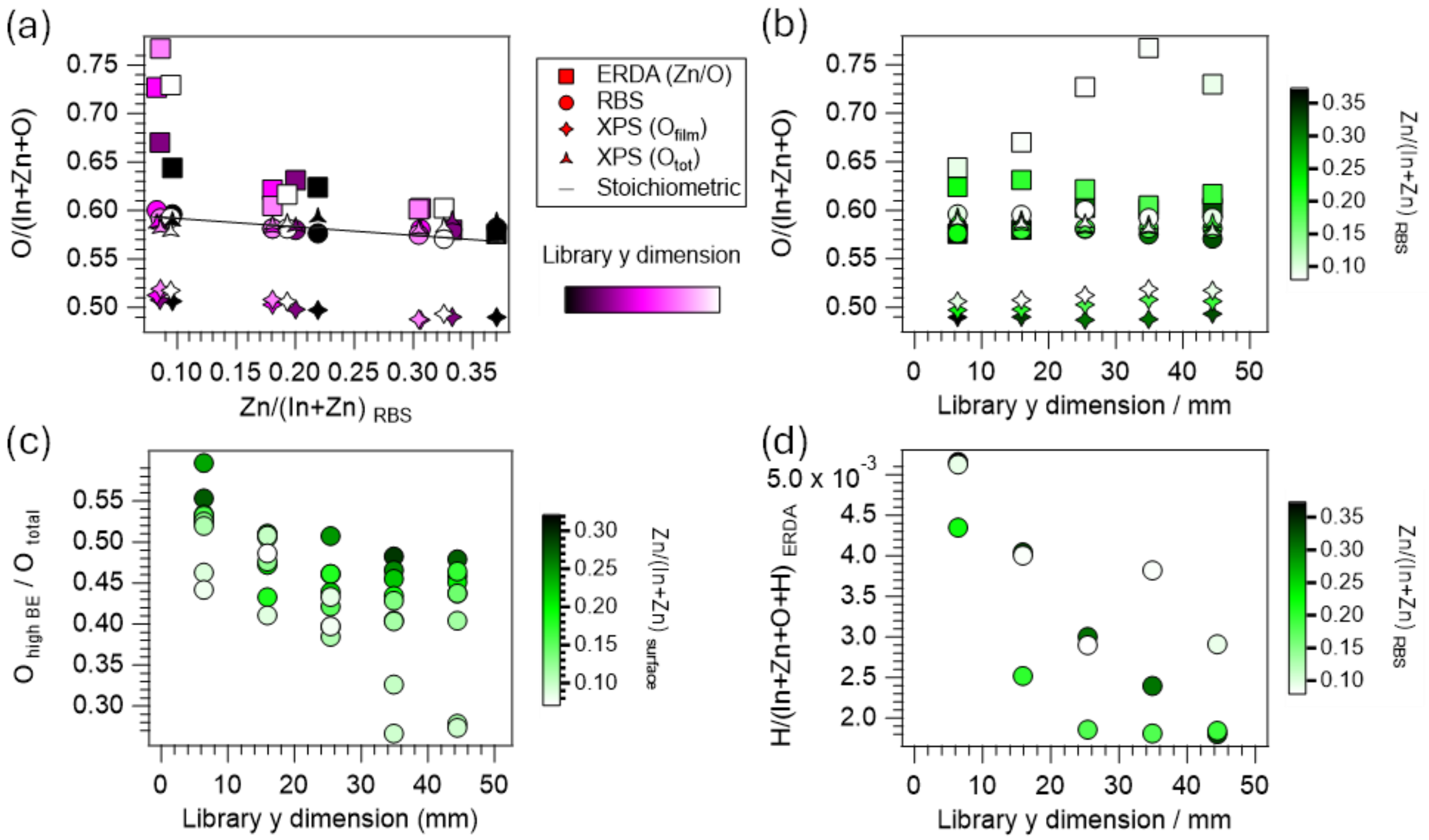

**Figure S5:** Oxygen concentrations as determined from ERDA, RBS and XPS after air-exposure as a function of (a) Zn/(In+Zn) content and (b) the library position. For the calculation of the ERDA results, the In/Zn ratio from RBS was used. The outliers in oxygen content for low Zn/(In+Zn) ratios determined by ERDA might be related to the narrow dimensions of the respective library strip, so that the beam hit its edge. Consequently, the oxygen measurement was influenced by a parasitic signal originating from the underlaying substrate. For XPS, two values are presented: $O_{tot}$ includes all oxygen components on the surface, while $O_{film}$ considers only the fitted metal oxide species. Variation of (c)

the ratio of the area of the high binding energy O 1s component relative to the total O 1s area of an air-exposed library in XPS and (d) hydrogen content as determined from ERDA as a function of the position on the library. Considering the estimated error of the ERDA and RBS results and applying Gaussian error propagation, the relative error for the H/(In+Zn+O+H) ratios amounts to ~10%. The UHV transferred sample presented in the main text in Fig. 2 differs from the one investigated with RBS/ERDA.

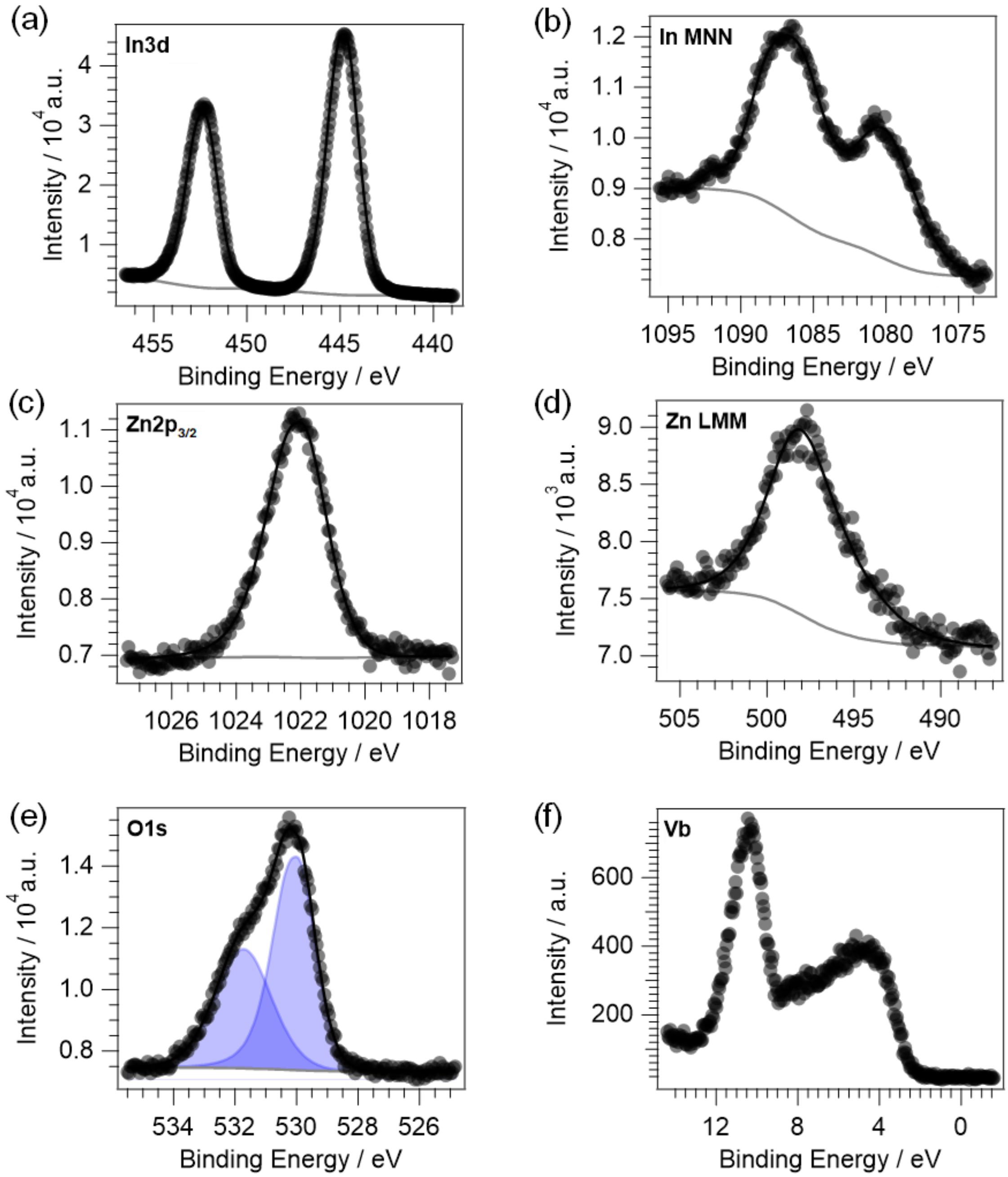

**Figure S6:** Typical XPS high-resolution spectra after air exposure: (a) In 3d, (b) In MNN, (c) Zn $2p_{3/2}$, (d) Zn LMM, (e) O 1 and (f) valence band with Shirley backgrounds. The O 1s spectrum clearly exhibits two species: The metal oxide peak at lower binding energies and an adsorbate peak. The metal oxide peak was fitted with an asymmetric peak shape (GL(40)T(2.1) in the CasaXPS software), while a symmetric line shape (GL(40)) was employed for the adsorbates. The asymmetric line shape was chosen based on the O 1s line shape after gentle sputtering as well as after a UHV transfer (c.f. **Figure S7**). The fit was constrained to the same FWHM to avoid varying widths among fits for different samples. Without this constraint, the absolute values of the relative areas vary, but the general trends in the adsorbate to metal oxide ratio and the $O_{film}$/(In+Zn+$O_{film}$) remain similar.

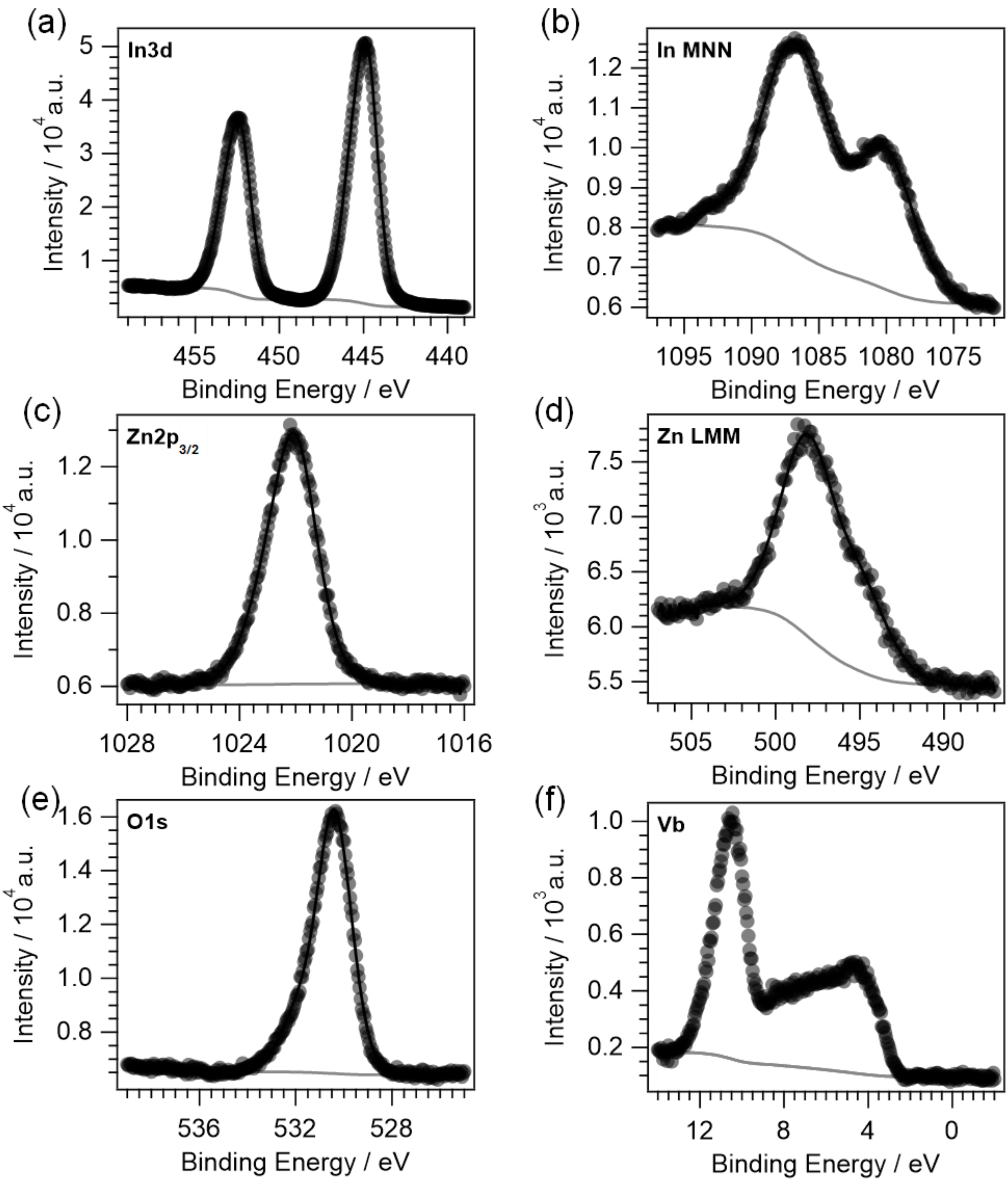

**Figure S7:** Typical XPS high-resolution spectra after UHV transfer (max. pressure $8{\cdot}10^{-8}$ mbar): (a) In 3d, (b) In MNN, (c) Zn $2p_{3/2}$, (d) Zn LMM, (e) O 1 and (f) valence band with Shirley backgrounds. The O 1s does not exhibit a pronounced second component but shows – as the In 3d and the Zn $2p_{3/2}$ core levels – an asymmetry towards higher binding energies as expected for samples with many free charge carriers.

## 3.2 Optical, electrical and structural characterization

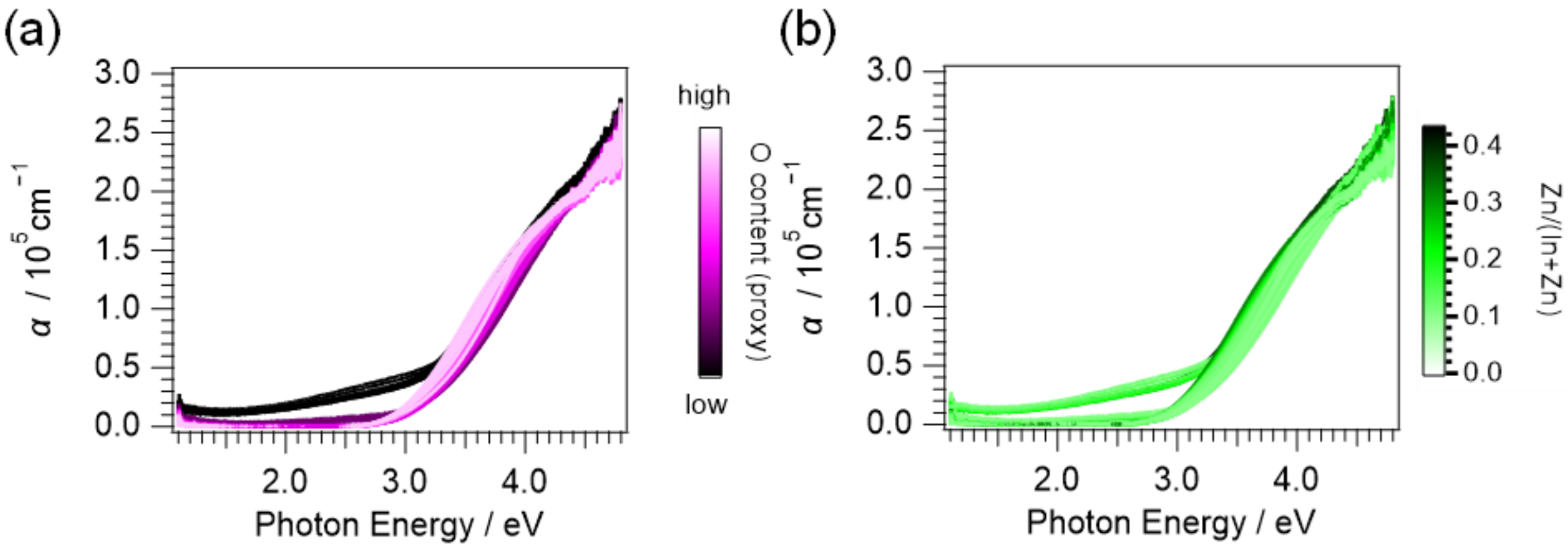

**Figure S8:** Absorption coefficient as determined from UV-vis spectrophotometry as a function of photon energy. The color codes represent (a) a proxy of the oxygen content and (b) the Zn/(In+Zn) ratio.

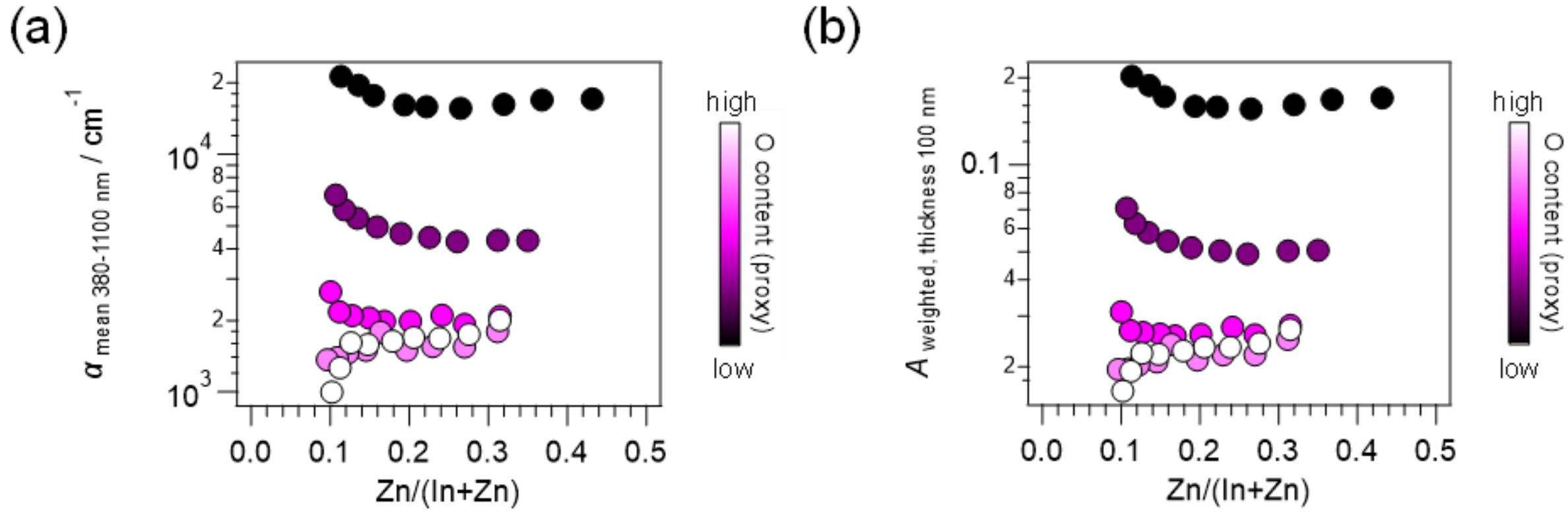

**Figure S9:** Absorption properties of an In-Zn-O-$v_O$ thin film library as a function of the Zn/(In+Zn) ratio with the oxygen content as a color code: (a) Absorption coefficient obtained from UV-vis mapping averaged between 380 and 1100 nm and (b) absorptance calculated from the absorption coefficients and a thickness of 100 nm, weighted by the AM1.5G spectrum and averaged between 380 and 1100 nm.

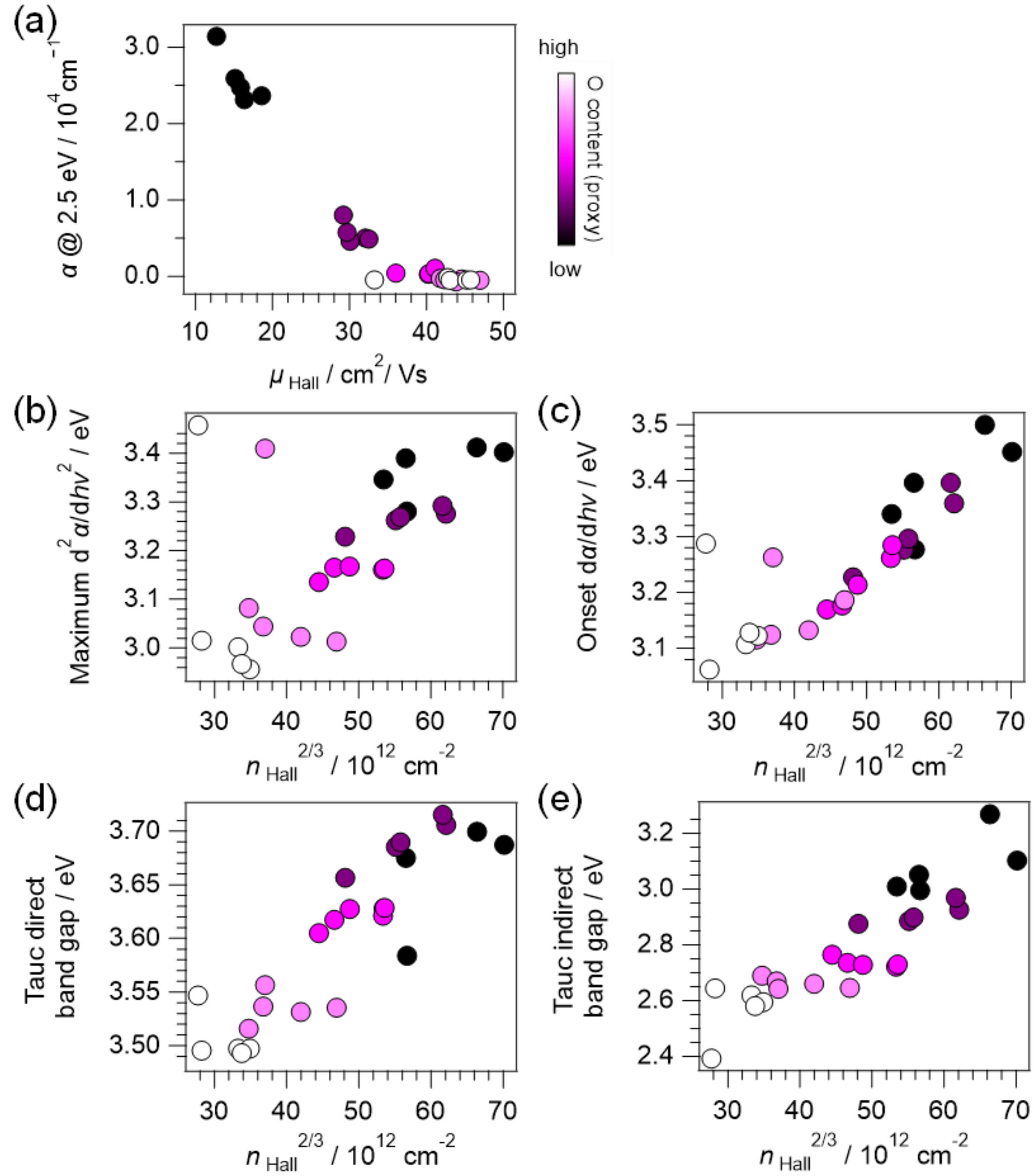

**Figure S10:** Correlation between electrical and optical properties: (a) Absorption coefficient below the band gap (at 2.5 eV) against the charge carrier mobility and (b-e) differently determined optical band gaps against the charge carrier concentration to the power of 2/3.

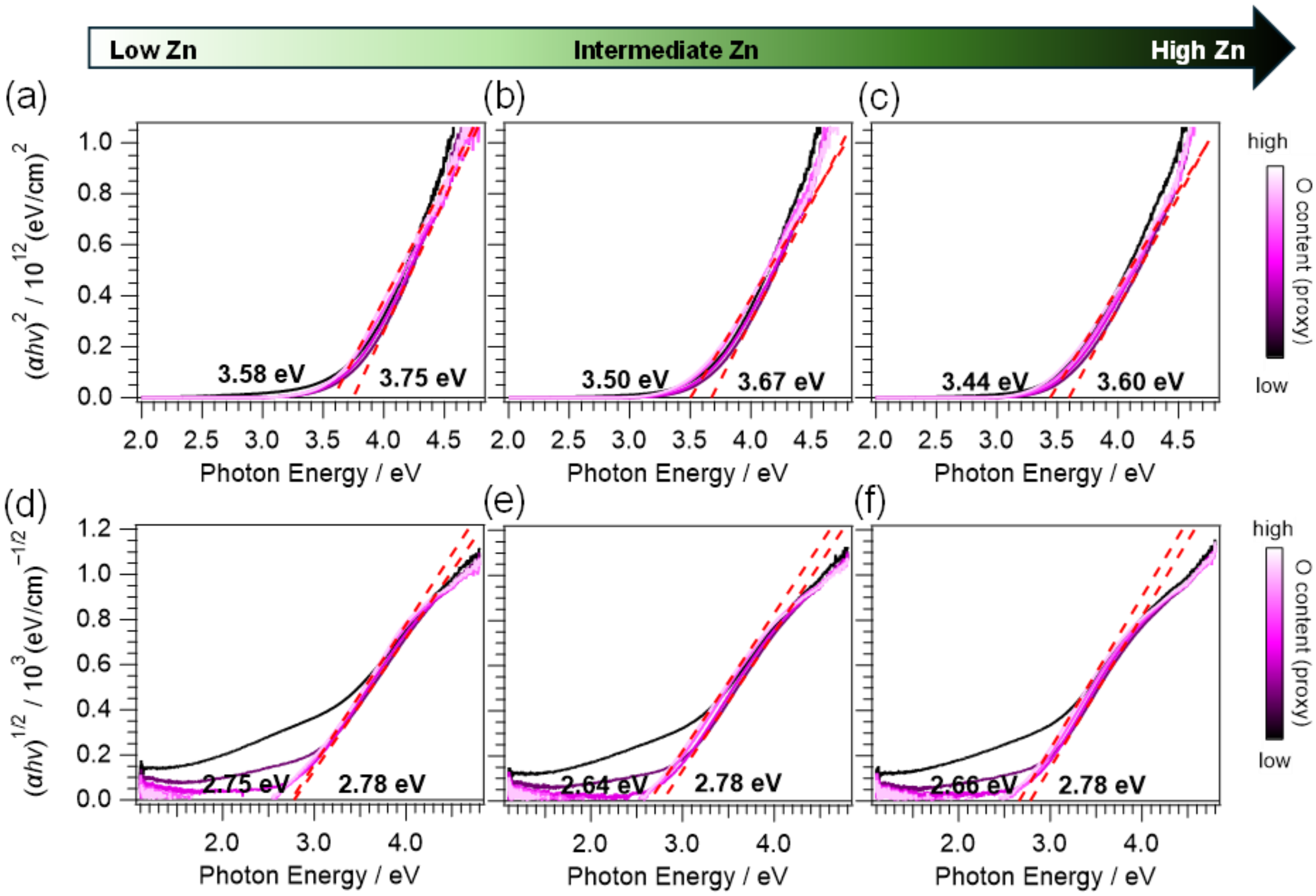

**Figure S11:** Tauc plots for the assumption of a direct band gap (a-c) and an indirect band gap semiconductor (d-f) with determined onset for exemplary compositions.

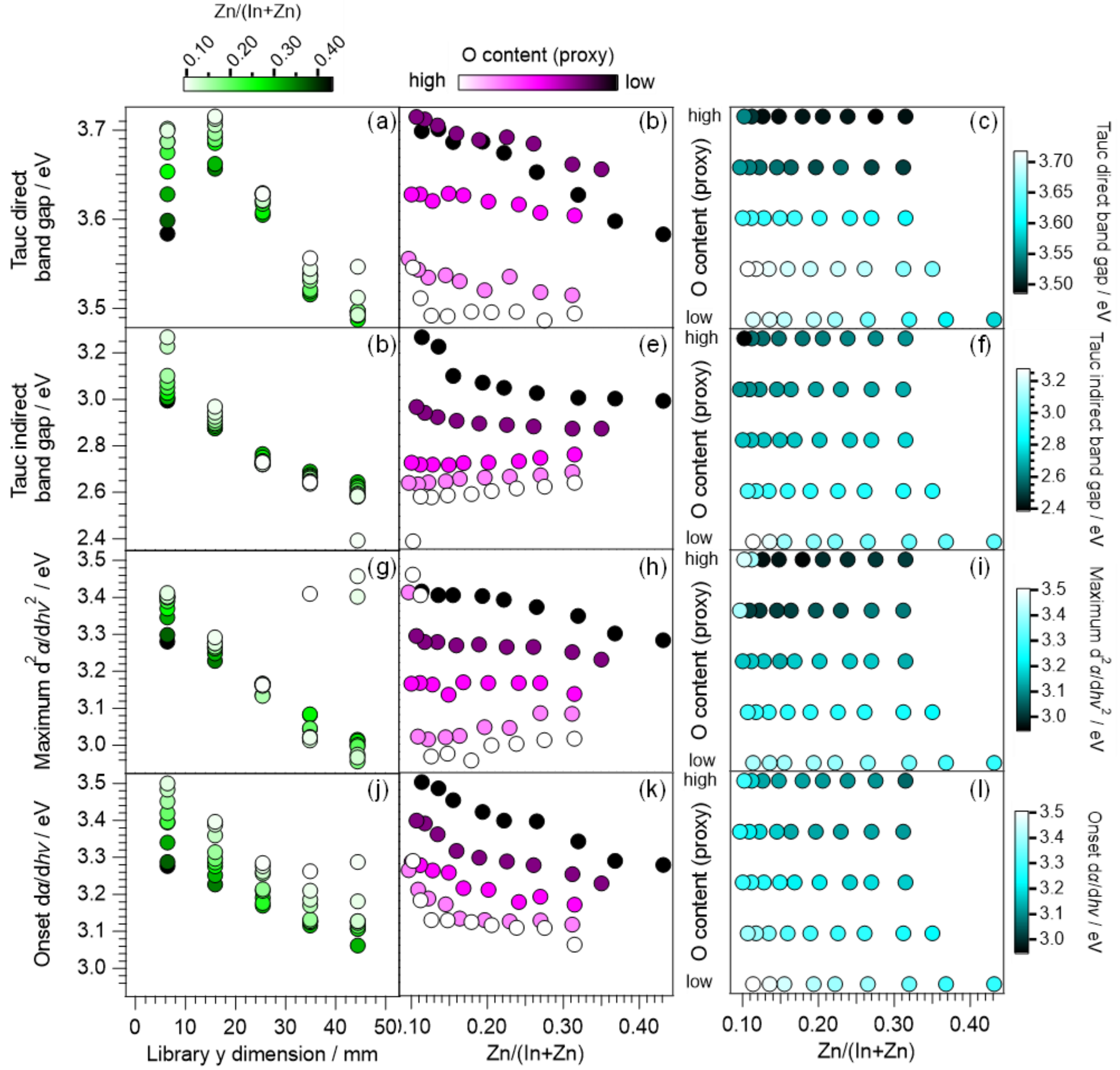

**Figure S12:** Optical band gaps of an In-Zn-O-$v_O$ thin film library determined by different methods as a function of the Zn/(In+Zn) ratio and the oxygen content: (a-c) direct and (d-f) indirect band gap based on Tauc Analysis using the code from Suram, Newhouse & Gregoire [1], (g-i) maximum in second derivative of the absorption coefficient and (j-l) onset of the derivative of the absorption coefficient against the photon energy using a threshold value of $10^5$ $cm^{-1}eV^{-1}$ (Comparable trends but varying absolute values are obtained by varying the threshold value (e.g. to $5 \cdot 10^4$ $cm^{-1}eV^{-1}$)),.

### 3.3 Cu diffusion barrier performance

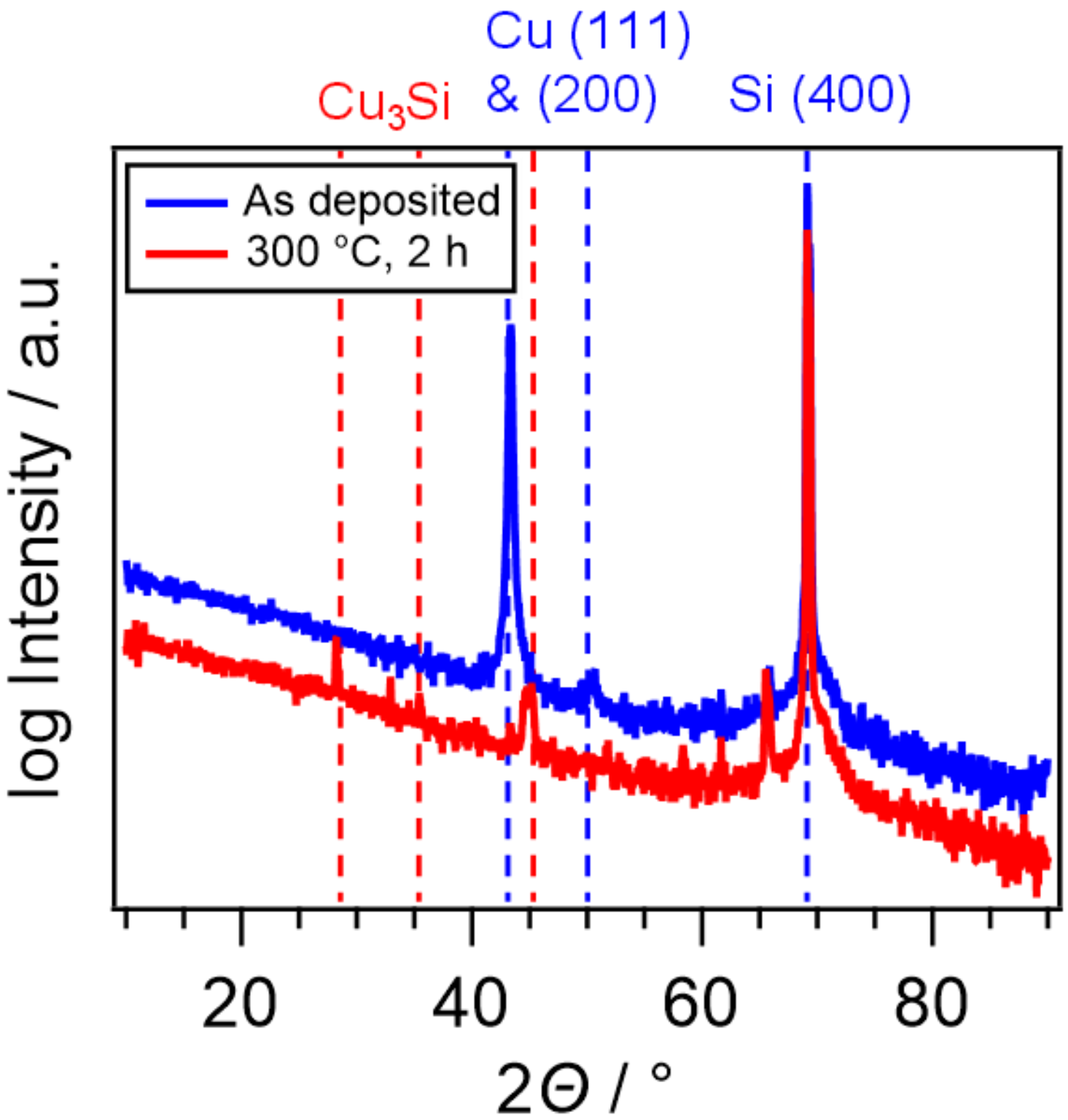

**Figure S13:** XRD pattern of a stack of Cu pads deposited at room temperature on a silicon wafer as deposited and after 2 h annealing at 300°C in Ar atmosphere. Dashed lines represent reflections extracted from literature references of fcc-Cu [2], Si [3] and $Cu_3Si$ [4].

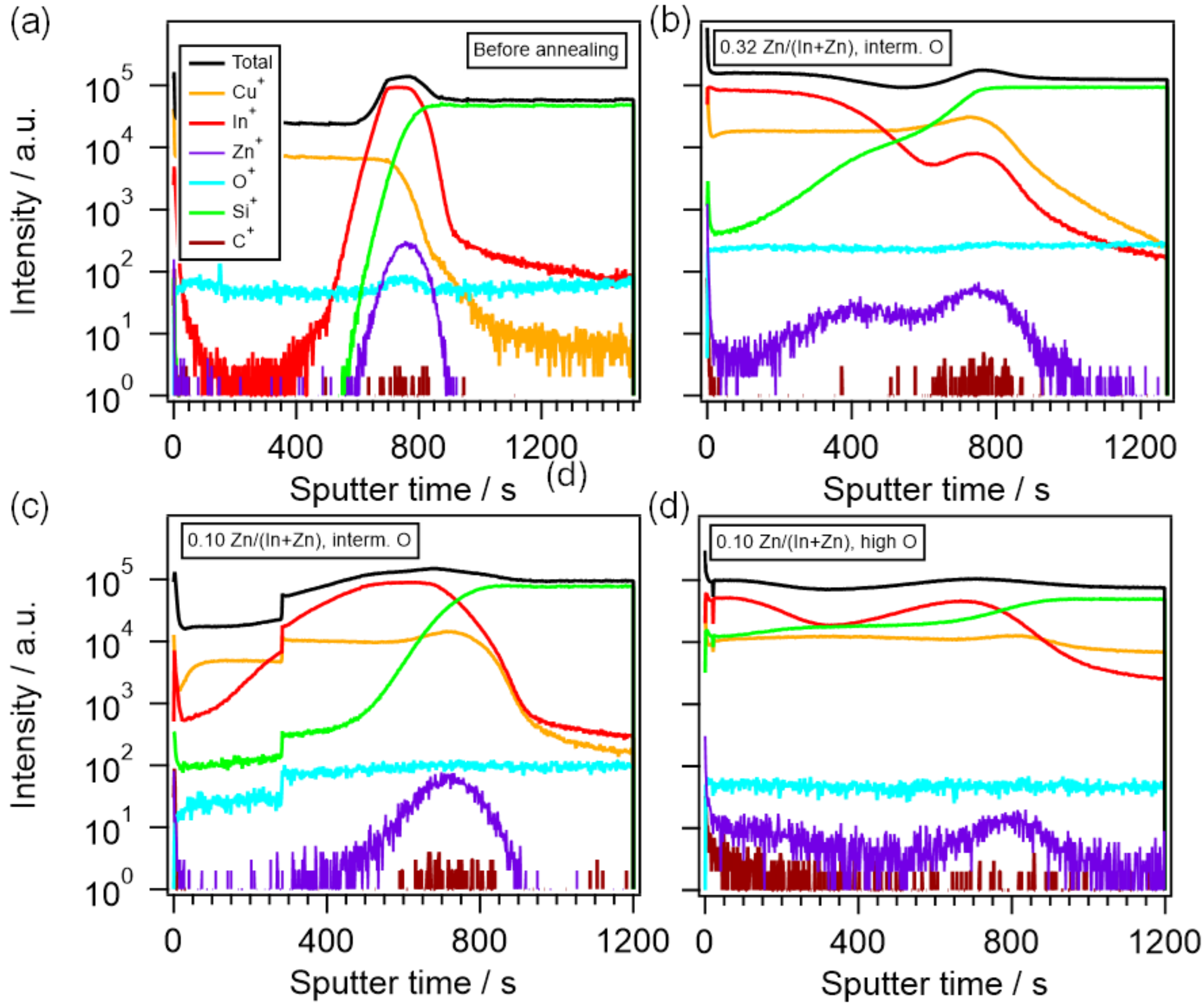

**Figure S14:** ToF-SIMS depth profiles performed with $O_2^+$ sputtering on a sample before annealing (a) and selected library samples after the annealing step of 450°C: Two samples with intermediate oxygen content and maximal (b) and minimal (c) screened Zn/(In+Zn) ratio, which both do not show Cu silicide formation in XRD; as well as a sample with high oxygen content and low Zn/(In+Zn) ratio, which exhibits copper silicide reflections in XRD (d). The discontinuity visible in all signals in (c) at around 300 s sputter time is related to a change in the LMIG current.

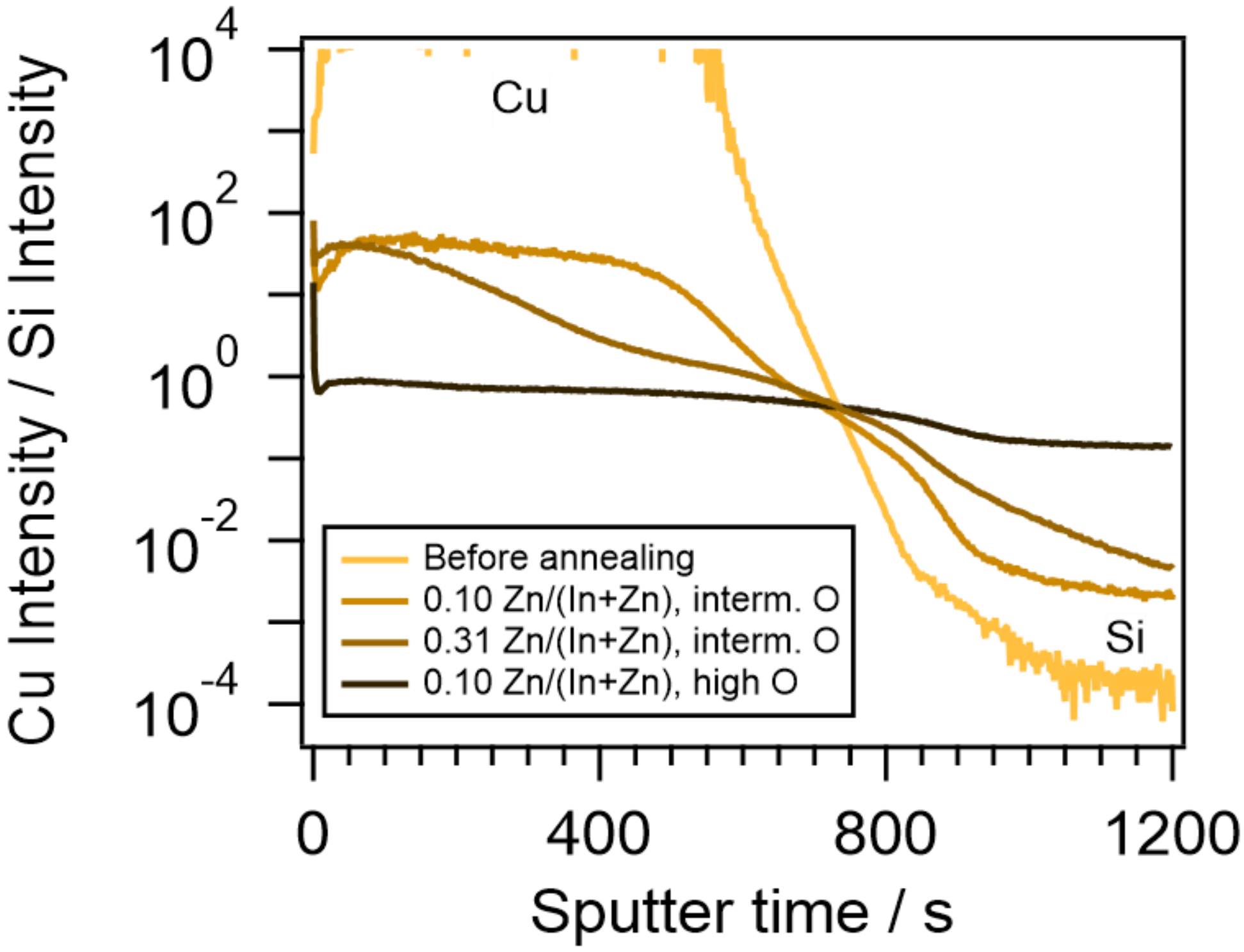

**Figure S15:** Ratio between Cu and Si signals extracted from the ToF-SIMS sputter depth profiles displayed in Figure S14. Cu silicides are detected exclusively in the sample with 0.10 Zn/(In+Zn) and high O contents.

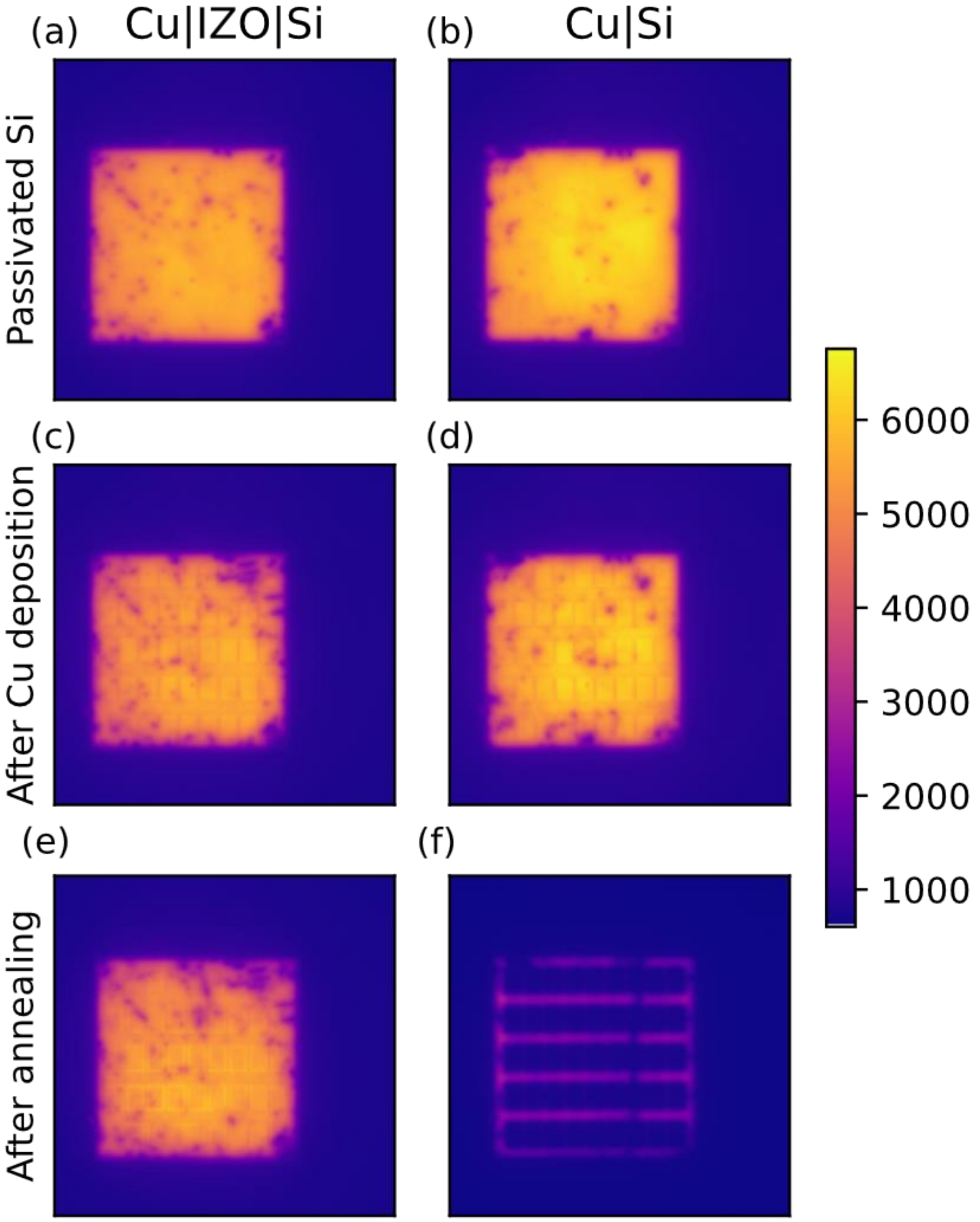

**Figure S16:** PL imaging on passivated Si wafers with (a,c,e) and without (b,d,f) the IZO library acting as a diffusion barrier for subsequently deposited copper pads: (a,b) Si wafers with poly-Si(n)|$SiO_x$ passivating contact stacks before IZO or Cu deposition, (c) after IZO and Cu deposition (Cu|IZO|Si), (d) respectively after Cu deposition (Cu|Si) and after annealing at 200°C in an Ar-filled glovebox for (e) 20 h (Cu|IZO|Si), (f) respectively for 1 h (Cu|Si). The color scales of all subfigures are unified. To assess the decay in PL after the different processing steps, the signal of circular areas with radius of 1 mm were integrated at the respective library positions after applying an edge detection algorithm.

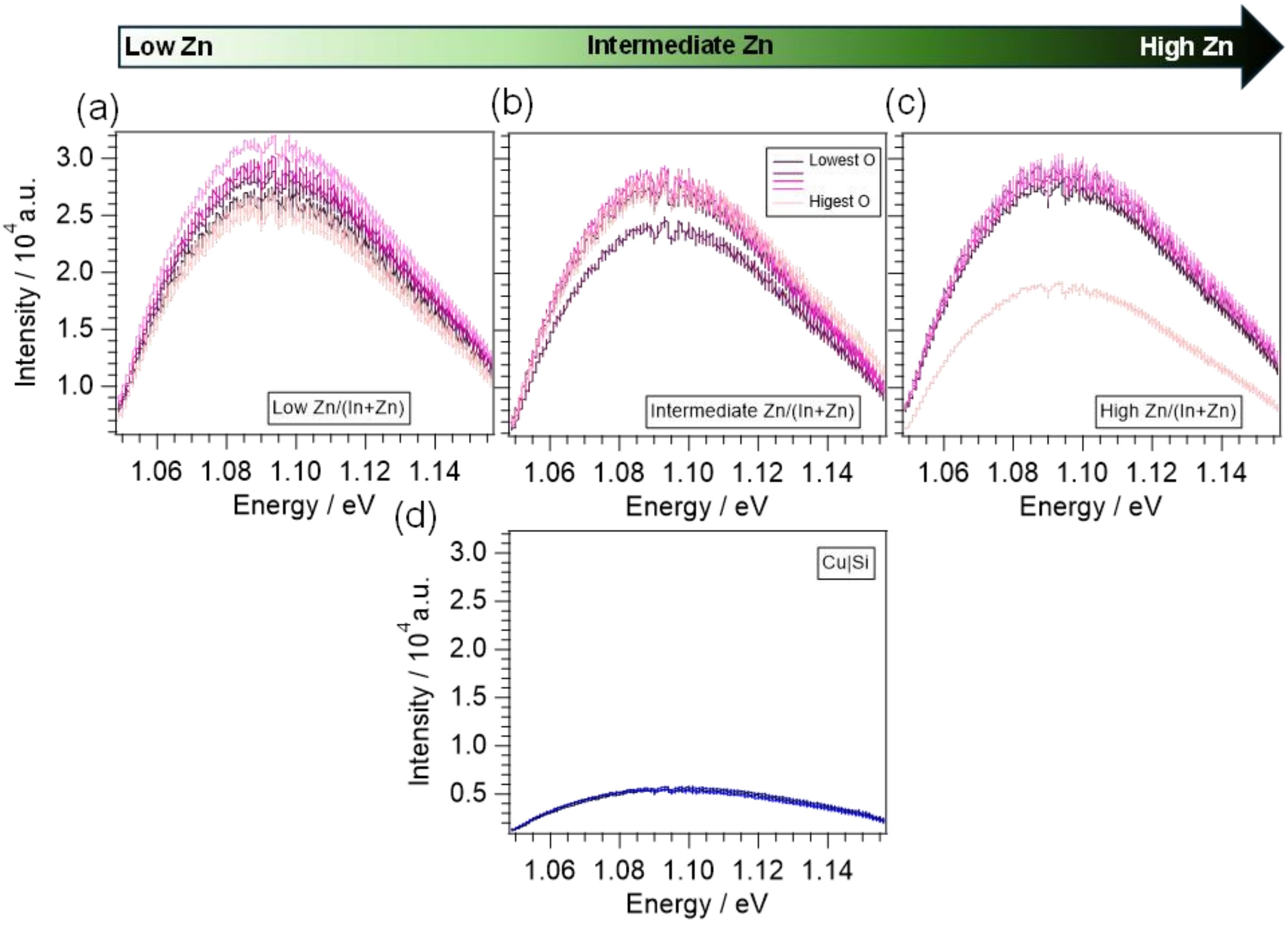

**Figure S17:** PL spectroscopy of the Si PL region at selected compositions of the Cu|IZO|Si library (a-c) after 20 h annealing at 200°C and the Cu|Si reference (d) after 1 h annealing at 200°C. The significantly lower PL intensity of the sample with highest oxygen content in (c) is related to the reduced PL signal in this corner of the passivated Si wafer due to the imprinted wafer ID.

## 3.4 UV stability

**Supplementary Note 1:**

The accelerated UV stability testing was performed by irradiation of an In-Zn-O-$v_O$ library by a UCUBE-10-305 UV source from UWAVE (305 nm, ~40 mW/cm$^2$) with a distance of 35 mm in air for 11 days. The respective UV dose (~105 kWh/ m$^2$) is estimated to an equivalent of 6 years of outdoor UV exposure following Ref. [5]. The UV aging was performed on an In-Zn-O-$v_O$ library on an EXG glass substrate deposited with a slightly higher oxygen gas flow than the those used for the Cu diffusion barrier testing. Therefore, the conductivity maximum is shifted slightly off the central row of the library, so that the regions of interest are shifted accordingly in Figure S18.

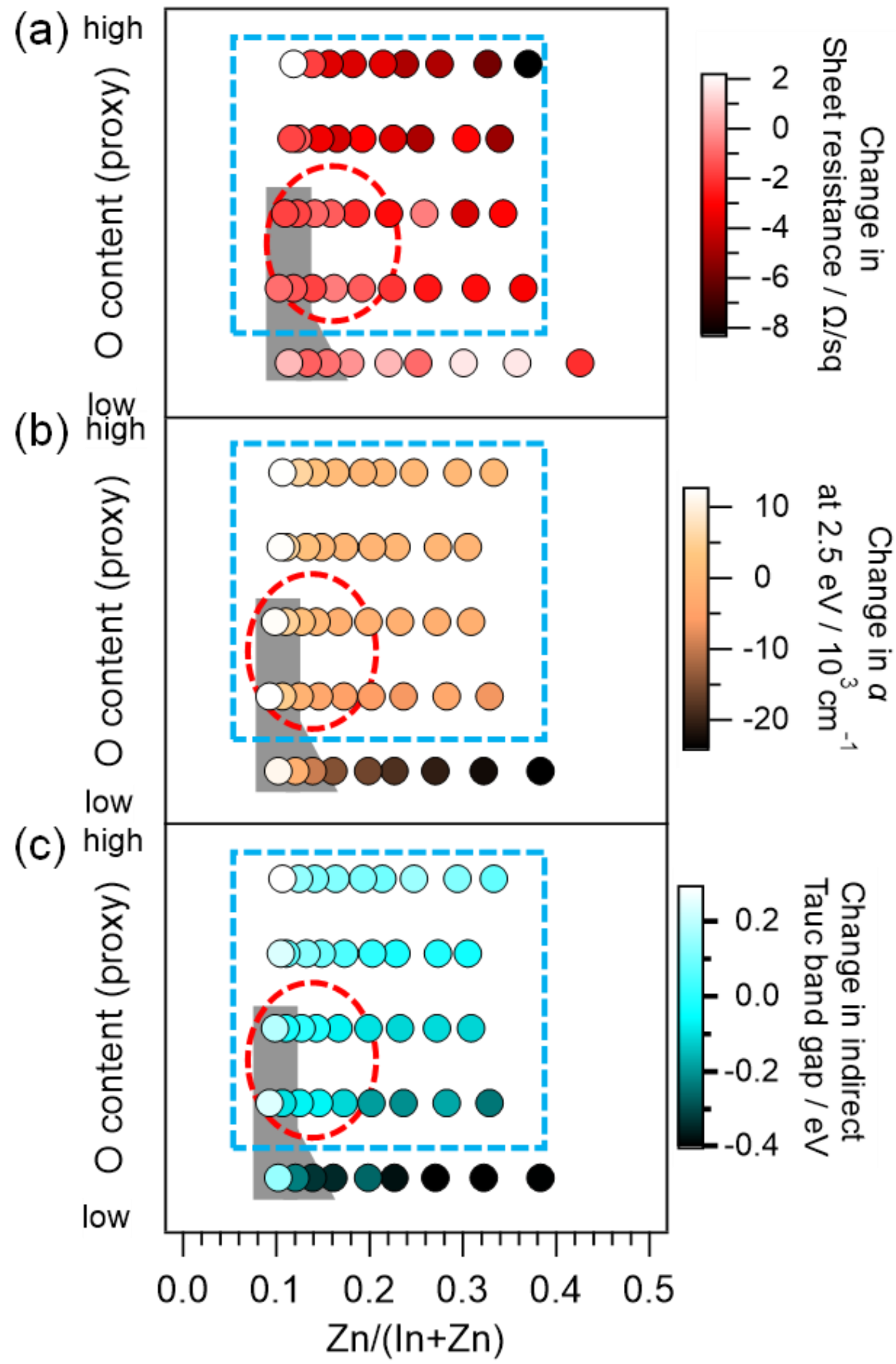

**Figure S18:** UV stability testing on an In-Zn-O-$v_O$ thin film library: Change in (a) sheet resistance, (b) sub-band gap absorption coefficient (at 2.5 eV), and (c) indirect optical gap by Tauc analysis after an equivalent dose of 6 years of outdoor UV exposure. The grey region indicates the presence of bixby-ite reflections in XRD scans, which did not alter significantly after UV exposure. The red dashed lines indicate the compositional region with conductivities above $2{\cdot}10^3$ S/cm and blue dashed lines regions with averaged absorption coefficients below $5{\cdot}10^3$ cm$^{-1}$.